\documentclass[12pt, draftclsnofoot, onecolumn]{IEEEtran}

\usepackage{slashbox}
\usepackage[T1]{fontenc}
\usepackage[utf8]{inputenc}
\usepackage[russian,english]{babel}
 \usepackage{siunitx}
 \usepackage{diagbox}
 \usepackage{multirow}
 \usepackage{vcell}
\usepackage{cuted}
\usepackage{flushend}
\usepackage{stfloats}
\usepackage[dvipsnames]{xcolor}
\usepackage{multirow}
\usepackage{caption}
\usepackage{bigints}
\usepackage{amssymb}
\usepackage[final]{graphicx}
\usepackage{amsmath}
\usepackage{url}
\usepackage{cite}
\usepackage{balance}
\usepackage{algorithm}
\usepackage{algorithmic}
\usepackage{graphicx}
\usepackage{amsmath}
\newcounter{hints}
\renewcommand{\thehints}{\roman{hints}}
\newcommand{\hintedrel}[2][]{%
  \refstepcounter{hints}%
  \if\relax\detokenize{#1}\relax\else\label{#1}\fi
  \mathrel{\overset{\mathrm{(\thehints)}}{\vphantom{\le}{#2}}}%
}

\usepackage[dvipsnames]{xcolor}
\definecolor{blue}{rgb}{0.00,0.00,1.00}
\definecolor{red}{rgb}{1.00,0.00,0.00}
\definecolor{yellow}{rgb}{0.00,1.00,0.00}
\DeclareMathOperator*{\argmin}{\arg\!\min}

\newcommand{\prf}{\textit{\textbf{Proof}:\;}}

\begin{document}

\title{Bit-Interleaved Coded Energy-Based Modulation with Iterative Decoding
\thanks{Ali~Fazeli and Ha H.~Nguyen are with the Department of Electrical and Computer Engineering, University of Saskatchewan, Saskatoon, Canada S7N 5A9. Emails: \{ali.fazeli, ha.nguyen\}@usask.ca.}
\thanks{Halim Yanikomeroglu is with the Department of Systems and Computer Engineering, Carleton University, Ottawa, Canada, K1S 5B6. Email: halim@sce.carleton.ca.}

\thanks{This work was supported in part by an NSERC Discovery Grant.}
}

\author{Ali~Fazeli, Ha H.~Nguyen, and Halim Yanikomeroglu}

\maketitle
\newtheorem{remark}{\textbf{Remark}}
\maketitle
\newtheorem{Convention}{\textbf{Convention}}
\maketitle
\newtheorem{definition}{\textbf{Definition}}
\maketitle
\newtheorem{lemma}{\textbf{Lemma}}
\maketitle
\newtheorem{theorem}{\textbf{Theorem}}
\maketitle
\newtheorem{corollary}{\textbf{Corollary}}
\maketitle
\newtheorem{example}{\textbf{Example}}

\vspace{-1.5cm}
\begin{abstract}
	This paper develops a low-complexity near-optimal non-coherent receiver for a multi-level energy-based coded modulation system. Inspired by the turbo processing principle, we incorporate the fundamentals of bit-interleaved coded modulation with iterative decoding (BICM-ID) into the proposed receiver design. The resulting system is called \emph{bit-interleaved coded energy-based modulation with iterative decoding (BICEM-ID)} and its error performance is analytically studied. Specifically, we derive upper bounds on the average pairwise error probability (PEP) of the non-coherent BICEM-ID system in the \emph{feedback-free} (FF) and \emph{error-free feedback} (EFF) scenarios. It is revealed that the definition of the \emph{nearest neighbors}, which is important in the performance analysis in the FF scenario, is very different from that in the coherent BICM-ID counterpart. The analysis also reveals how the mapping from coded bits to energy levels influences the diversity order and coding gain of the BICEM-ID systems. A design criterion for good mappings is then formulated and an algorithm is proposed to find a set of best mappings for BICEM-ID. Finally, simulation results corroborate the main analytical findings.
\end{abstract}

\IEEEpeerreviewmaketitle

\section{Introduction}
\IEEEPARstart{O}n-off keying (OOK) and frequency-shift keying (FSK) \cite{Book-Proakis} are the most basic forms of energy-based modulation (EBM) at radio frequency (RF). Fundamentally, since EBM encodes information only in the signal amplitude, it is completely immune to random phase changes introduced by the wireless channel on the received signal. This property eliminates the need for a channel estimator at the receiver, and enables the implementation of low-cost receivers based on \emph{non-coherent} detection. Another family of modulation that can work without channel estimation at the receiver is differential modulation. Although generally having lower spectral efficiencies than differential modulation, EBM is preferred or the only choice in many scenarios \cite{Torrieri04,Valenti05,Fabregas07,Choi18IM,Manolakos2016twc,Chowdhury16,Jing2016,Cuba19,Xie19,Xie2020}, some of which are discussed below.
\begin{itemize}
	\item Hardware impairments, such as in-phase/quadrature (I/Q) imbalance and phase noise (in local oscillators), are hard to mitigate. For example, such issues arise in low-power wireless systems such as sensor networks. These systems use EBM to dispense with the need for I/Q chains, high-quality local oscillators and mixers \cite{Manolakos2016twc}.

	\item The wireless channel changes significantly from symbol to symbol. This situation happens when the channel is very rapidly varying. Systems that employ fast frequency hopping, or operate under high velocity are examples of this scenario \cite{Torrieri04,Valenti05,Fabregas07,Choi18IM}. In these cases, neither coherent nor differential detection is a good solution.

	\item The system's complexity favors simple, robust designs. For example, the use of a very large antenna array at the receiver means a large number of RF chains. As such, simple signal processing of the signal received at each antenna is critical to reduce complexity and cost \cite{Chowdhury16,Jing2016,Manolakos2016twc,Cuba19,Xie19,Xie2020}.
\end{itemize}

The basic EBM schemes, such as OOK and FSK, use two energy levels, and hence suffer from a low spectral efficiency (SE). This drawback has motivated various studies on improving the SE of EBM schemes. A natural extension of OOK is $(M+1)$-ary amplitude-shift keying (ASK), whose signal constellation consists of $(M+1)$ equidistant signal points, where $M\geq 1$, and each signal point can carry $m=\log_2 (M+1)$ bits. Unfortunately, ASK has unacceptably poor performance when the channel experiences Rayleigh fading \cite{Manolakos2016twc}. The problem of constellation design for multi-level EBM is considered in
\cite{Manolakos2016twc} for a massive single-input multiple-output (SIMO) system. In particular, the authors present algorithms for obtaining optimal energy levels to maximize the error exponent of the symbol error probability (SEP) with respect to the number of receiver antennas. A similar work is carried out in \cite{Jing2016}, but for a massive multiple-input multiple-output (MIMO) system. It is pointed out that the results in \cite{Manolakos2016twc,Jing2016} are obtained based on asymptotic approximations in the limit of very large number of antennas. In contrast, the authors in \cite{Gao2020} consider the problem of optimal constellation design (energy levels) for a non-coherent multi-level EBM system by minimizing the \emph{exact} SEP subject to a power constraint. In other words, the work in \cite{Gao2020} addresses the problem of finding optimal energy levels for an EBM system equipped with an \emph{arbitrary} number (not necessarily a massive number) of receive antennas, at any signal-to-noise ratio (SNR), and for any modulation order.

All the aforementioned works on multi-level EBM are restricted to \emph{single-shot} modulation, i.e., the case that one signal point is transmitted over channel and demodulated correspondingly at the receiver. Unfortunately, even with the optimal constellation design, the error performance of single-shot EBM systems degrades drastically when the number of \emph{positive} energy levels is larger than one ($M\geq 2$). To address this issue, a framework of non-coherent multi-level EBM has recently been theoretically explored in the context of index modulation (IM) in \cite{Fazeli22}. The present work is a sequel to \cite{Fazeli22}, which investigates practical aspects of \cite{Fazeli22}. To clearly clarify the contributions of our work, a brief overview of IM techniques is given next.

\subsection{Overview of Index Modulation}

The term ``index modulation'' refers to a wide range of techniques that transmit a portion (or all) of information bits through patterns of active and inactive indices. These indices can be for transmit antennas (in MIMO systems), sub-carriers (in OFDM systems), time slots, etc., or a combination of them. In a typical IM system, the input bit stream is divided into two sub-streams. Bits in the first sub-stream determine which indices to be activated, and bits in the second sub-stream modulate the activated indices with either a one-dimensional (e.g., ASK) or two-dimensional (e.g., PSK or QAM) constellation. Furthermore, IM techniques can be classified in different ways, two of which are discussed in the following \cite{Basarjul16IM,Ishikawa18,Mia_Wen_2019}.
\begin{itemize}
	\item \emph{Nature of index:} Based on the available resources, various types of IM exist. For instance, a MIMO system can use indices of the transmit antennas for transmission in the index domain. Techniques such as spatial modulation (SM) and space-shift keying (SSK) fall in this type of IM \cite{Mesleh2008,Jeganathan2009}. As another example, when indices of the sub-carriers are used in an OFDM-based system, the resulting system is OFDM-IM \cite{Basar2013}.

	\item \emph{Coherent or non-coherent:} Similar to many modulation techniques, IM can be designed to work with a coherent or non-coherent receiver \cite{Chang2012,Basar2013,Basar2015,Miao_wen2017}, depending on the availability of the channel state information (CSI) and/or the affordable implementation complexity.

\end{itemize}

While the difference and applicability of coherent and non-coherent systems were discussed before, in the following we elaborate on non-coherent IM techniques since the focus of this paper is on non-coherent communication.

There are generally two approaches to realize an IM scheme without requiring CSI, each having its own advantages and applications: \emph{differential IM} and \emph{energy-based (EB) IM}. Differential space-time shift keying (DSTSK) is one of the first works on differential IM \cite{Sugiura2010}. In this technique, communication in the index domain is performed by \emph{indexing dispersion matrices} that belong to a group of pre-designed space-time matrices. The work in \cite{Sugiura2010} is limited to ASK, and generalized to PSK and QAM in \cite{Xu2011siglet,Sugiura2011}. Later, the authors in \cite{Bian13,Bian2015} developed differential SM (DSM), a non-coherent counterpart of the original SM \cite{Mesleh2008}. Since early DSM systems utilize differentially-encoded unitary matrices, they suffer the lack of scalability. This issue was later addressed in \cite{Ishikawa2017,Ishikawa2018tcom,Xiao2019}, which develops rectangular DSM. As research on differential IM schemes is vast, the interested reader is referred to \cite{Basar17,Ishikawa18,XU19} for more complete surveys of recent advances.

When information is transmitted only in the index domain, an EBM system is resulted \cite{Choi18IM,Gopi19,Fazeli2019,Fazeli2019Letter,Nguyen2019}. To facilitate the discussion, we shall describe the transmitter by a $(V,L)$-code. This means that the transmitter accepts a binary vector of length $L$ as its input and generates a length-$V$ vector of symbols as its output. This output vector is carried by $V$ available sub-channels (indices). The SE of such a system is $\frac{L}{V}$ (bits/channel use). For the non-coherent IM system in \cite{Choi18IM}, $V$ sub-channels (indices) are grouped into $G$ clusters, each having $V_1 = V/G$ sub-channels. In each cluster, the activation patterns are designed using a maximum-size constant-weight binary code of length $V_1$ and Hamming weight $Q$. Therefore, the designed system has a SE of $\frac{\left\lfloor \log_2\binom{V_1}{Q}\right\rfloor}{V_1}$. The main idea in \cite{Choi18IM} is extended in \cite{Gopi19,Fazeli2019,Fazeli2019Letter,Nguyen2019} to yield better SE and/or higher diversity order. Specifically, in \cite{Gopi19}, in addition to sub-channels in each cluster, the $G$ clusters introduced in \cite{Choi18IM} are also indexed. Hence, by adding a second index domain, which selects $K$ out of $G$ clusters, the system's SE increases to $\frac{\left\lfloor\log_2\binom{G}{K}\right\rfloor}{GV_1}$. On the other hand, the work in \cite{Fazeli2019} generalizes the system in \cite{Choi18IM} by employing codes with a more flexible structure (binary codes that are neither constant-weight nor maximum-size) for realizing activation patterns in each cluster. The weight flexibility of activation patterns helps to increase the SE to its maximum value of $1$, which is higher than the SE of the systems in \cite{Choi18IM} and \cite{Gopi19}. In \cite{Fazeli2019Letter} the authors consider the
code design problem for the system proposed in \cite{Fazeli2019} and showed that it is equivalent to the problem of code design for an ideal binary asymmetric channel. While the analysis in \cite{Fazeli2019Letter} mainly focuses on the case in which sub-channels are independent, the analysis in \cite{Nguyen2019} examines the case where all sub-channels experience the same fading, a situation that arises in narrow-band chirp spread spectrum communications.

\subsection{Contributions}

\begin{figure}[t!]
	\centering
	\includegraphics[width=3.5in]{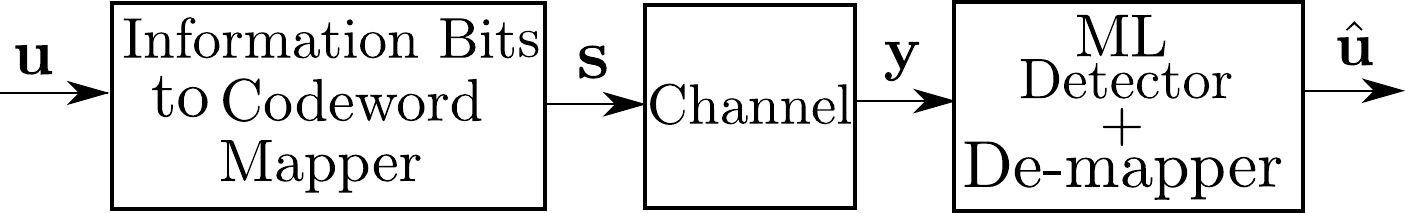}
	\caption{Block diagram of a non-coherent multi-level IM system \cite{Fazeli22}.}	 \label{fig_general_diag}
\end{figure}

Although the EB IM schemes suggested in \cite{Gopi19,Fazeli2019,Fazeli2019Letter,Nguyen2019} enjoy good error performance, their spectral efficiencies are at most 1 (bit/channel use) because they are based on binary (two-level) activation patterns. More recently, the concept of single-shot non-coherent multi-level EBM was merged with EB IM, which results in a system called \emph{non-coherent multi-level IM}, and depicted in Fig. \ref{fig_general_diag}. As seen in the figure, a binary vector $\mathbf{u}$ of length $L$ is mapped to a codeword $\mathbf{s}$ in a $(V,L)$ code-book. The code-book is constructed from $M+1$ alphabets, which are square roots of asymptotically-optimal energy levels as designed in \cite{Gao2020} for single-shot non-coherent $(M+1)$-level EBM. The codeword $\mathbf{s}$ is transmitted over $V$ flat fading sub-channels. Finally, at the receiver, an estimate $\hat{\mathbf{u}}$ of $\mathbf{u}$ is obtained by, for example, the maximum likelihood (ML) sequence detector, followed by a look-up table de-mapper. The superiority of such a \emph{multi-level} scheme over the single-shot multi-level EBM in \cite{Chowdhury16,Jing2016,Manolakos2016twc,Cuba19,Xie19,Xie2020} and EB two-level IM in \cite{Gopi19,Fazeli2019,Fazeli2019Letter,Nguyen2019} in terms of error rate and SE is theoretically proved in \cite{Fazeli22}.

While \cite{Fazeli22} provides theoretical analysis to demonstrate the advantages of the non-coherent multi-level IM system in Fig.~\ref{fig_general_diag}, it lacks details on practical realization and implementation of such a system. Specifically, the analysis in \cite{Fazeli22} is carried out without specifying the exact operation of the block ``Information Bits to Codeword Mapper'' in Fig. \ref{fig_general_diag}, and by assuming the optimal ML receiver. This paper examines practical design aspects of the system in Fig.~\ref{fig_general_diag} and makes the following contributions.
\begin{itemize}
\item Presenting a practical realization of the transmitter in Fig. \ref{fig_general_diag} that is inspired by the technique of bit-interleaved coded modulation (BICM) \cite{Caire98}. Specifically, the relation between the binary vector $\mathbf{u}$ and the multi-level vector $\mathbf{s}$ is specified through the concatenation of an encoder (such as a  binary convolutional encoder), a random interleaver, and a multi-level EB modulator. As such, the transmission method is naturally called bit-interleaved coded EB modulation (BICEM) and has the following key advantages: (i) having a simple design, as the system enjoys a \emph{fully non-coherent receiver, which bypasses the need for accurate channel estimators}, (ii) exploiting the inherent diversity in fading channels, and (iii) facilitating the implementation of a low-complexity suboptimal iterative receiver whose performance closely approaches that of the optimal receiver \cite{PhD_Tran08,Chindapol01,Tran06-Broad}.

\item Developing a low-complexity suboptimal iterative receiver, which performs turbo processing between the soft-output EB demodulator and the soft-input soft-output (SISO) decoder that are linked by interleaver/deinterleaver. It is stressed that the EB modulator/demodulator employed in the structure of BICEM-ID facilitate reliable communication without requiring CSI at the receiver.

\item Performing analysis on the average pairwise error probability (PEP) of the system in two scenarios: \emph{feedback-free} (FF) (i.e., there is no feedback from the decoder to the demodulator), and \emph{error-free feedback} (EFF) (i.e., when there is no error in the feedback from the decoder to the demodulator). The average PEP analysis leads to a design criterion for minimizing the error rate of the system and also shows how the system's diversity order and coding gain depend on design factors such as signal mapping, modulation order, minimum Hamming distance of the code, and the number of receive antennas.

	\item Developing a general algorithm for finding a set of best mappings for the proposed system. Performance of an $8$-ary BICEM-ID system is investigated in detail under different signal mappings.
	
\end{itemize}

%

	

The applications and advantages of non-coherent EBM over coherent (and differential non-coherent) communication were extensively studied. Before closing this section, it is worth comparing the structure of BICEM-ID and traditional BICM-ID to understand what distinguishes our proposed scheme from existing works such as \cite{Chindapol01,Li02,Tran06-Broad,Valenti05,Wei15,Baeza18}. In a nutshell, all new analytical results/findings concerning BICEM-ID stems from the change in the structure of modulator and SISO demodulator. The majority of works on BICM-ID, e.g., \cite{Chindapol01,Li02,Tran06-Broad}, focus on coherent communication. In such cases, the modulator converts bits into phase-modulated signals, and the demodulator's output is a function of the CSI. There also exist a few works such as \cite{Valenti05,Wei15,Baeza18} that focus on non-coherent BICM-ID. Yet, they either employ binary EB modulators \cite{Valenti05} or differential phase modulators \cite{Wei15,Baeza18}. In contrast to [3], BICEM-ID should deliver a high SE, hence the modulator generates more than two energy levels. This imposes a unique constraint on constellation design, namely \emph{the geometric distancing} of signal points. The BICEM-ID's modified modulator/demodulator is the base of a number of new results, distinguishing it from the aforementioned works, substantially. For instance, analytical results show that the definition of nearest neighbor needs to be revised. Furthermore, it is shown that BICEM-ID is different from its traditional counterparts \cite{Chindapol01,Li02,Tran06-Broad} regarding  the impact of signal mapping on system performance. Hence, the problem of good signal mapping for the proposed system needs to be reconsidered.

The remainder of the paper is organized as follows. Section \ref{sec-sys} presents the system model of the proposed BICEM-ID and develops a soft-output demodulator for the proposed system. Performance analysis and design criterion are included in Section \ref{sec-ana}. An algorithm to find the best signal mappings for the proposed system is described in Section \ref{sec-alg}. Section \ref{sec-num} provides simulation results to illustrate the advantages of the BICEM-ID system. Finally, Section \ref{sec-con} concludes the paper.

\emph{Notations:} Throughout this paper boldface letters are used to denote column vectors; The $v$th element of a vector $\mathbf{s}$, is represented by $[\mathbf{s}]_v$; underlining is used to distinguish matrices from vectors; the $i$th column of a matrix $\underline{\mathbf{m}}$ is denoted by $[\underline{\mathbf{m}}]_{:,i}$; ${\mathcal C}{\mathcal N} (\mathbf{0},\mathbf{I}_V)$ refers to a circularly symmetric complex Gaussian vector with mean $\mathbf{0}$, and identity covariance  matrix $\mathbf{I}_V$ of size $V$;  diag($\mathbf{v}$) denotes a diagonal matrix with elements of $\mathbf{v}$ on the main diagonal; for integers $k$ and $m$, the quotient and remainder of the division $\frac{k}{m}$ are shown by $k\setminus m$, and ${\rm rem}(k,m)$, respectively; $\mathbf{s}^{\top}$ is the transpose of $\mathbf{s}$.

\section{System Model}\label{sec-sys}

\begin{figure}[t!]
	\centering
	\includegraphics[width=4.5in]{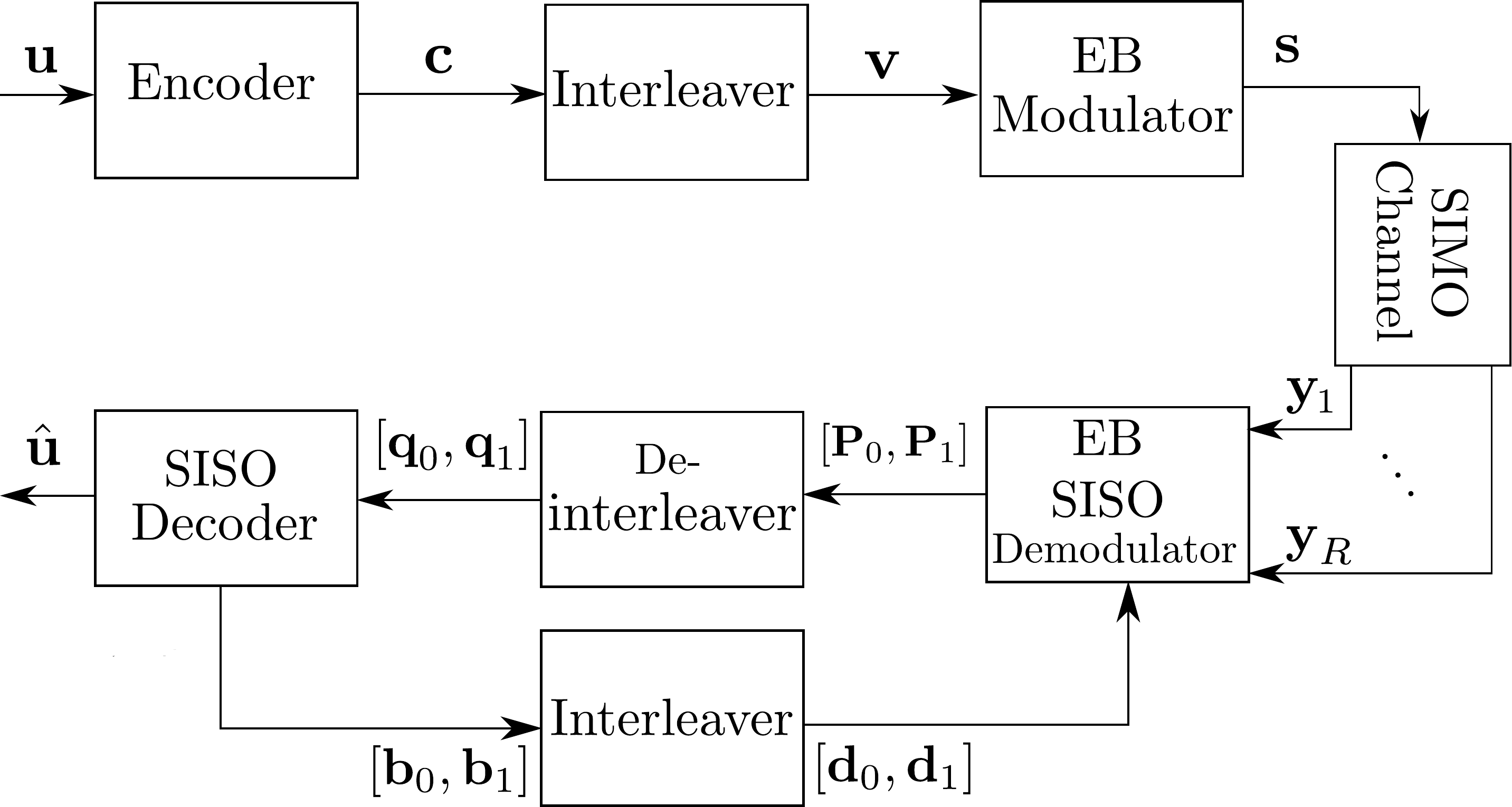}
	\caption{Block diagram of the proposed BICEM-ID system with an iterative receiver.}
	\label{fig_block_BICM_ID}
\end{figure}

Figure \ref{fig_block_BICM_ID} depicts a block diagram of the proposed BICEM-ID system. The single-antenna transmitter is a serial concatenation of a $(n_{\rm c},k_{\rm c},\nu_{\rm c})$ convolutional encoder, a bit interleaver, and a memoryless EB modulator. Each block of $L_{\rm I}$ information bits is appended with $\nu_{\rm c}$ termination bits to form an input binary vector $\mathbf{u}$ of length $L=L_{\rm I}+\nu_{\rm c}$. The encoder generates a binary vector $\mathbf{c}$ of $L_{\rm c}=\frac{n_{\rm c}L}{k_{\rm c}}$ coded bits. To break correlation of fading affecting coded bits, the vector $\mathbf{c}$ passes through a random interleaver. The modulator can be described by a one-to-one mapping $\xi: \{0,1\}^m \rightarrow \Psi$, where the constellation $\Psi$ is defined as $\Psi=\{\sqrt{p_0},\sqrt{p_1},\cdots,\sqrt{p_M}\}$, with $M=2^m -1$. Thus, for each group of $m=\log_2(M+1)$ coded bits $(b_1,\cdots,b_m)$, the modulator produces a signal $s=\xi(b_1,\cdots,b_m)\in\Psi$.

The design of an asymptotically-optimal constellation for non-coherent multilevel EBM has been addressed in \cite{Gao2020}. Specifically, let $E_{\rm s} =  \frac{1}{M+1}\sum_{l=1}^{M} p_l$ be the average energy of the constellation, $N_0$ the one-sided power spectral density of white Gaussian noise, and $\gamma = \frac{E_{\rm s}}{N_0}$ the signal-to-noise ratio (SNR). Then, the $(l+1)$th signal point in $\Psi$ is given by
\begin{IEEEeqnarray}{ll}\label{Eq_singleshot_levels}
	\sqrt{p_{l}}=\sqrt{(r^{l}-1)N_0}, \ \ \ \ \ 0\leq l \leq M,
	\IEEEeqnarraynumspace
\end{IEEEeqnarray}
where $r$ is the solution to the following equation \cite{Gao2020}:
\begin{IEEEeqnarray}{ll}\label{Eq_all_one_poly}
	\sum_{l=0}^{M}r^l-(M+1)(\gamma+1)=0.
	\IEEEeqnarraynumspace
\end{IEEEeqnarray}
The output $\mathbf{s}$ of the modulator is a vector of $V=\frac{L_{\rm c}}{m}$ symbols, with each symbol ($[\mathbf{s}]_v$, $v=1,\cdots,V$) taking a value in $\Psi$.

With $R$ antennas, the receiver can collect $R$ versions of the transmitted vector $\mathbf{s}$ that are disturbed by fading and noise. Specifically, the signal received at the $a$th antenna can be written as
\begin{IEEEeqnarray}{c}
	\label{Eq_received_sig}
	\mathbf{y}_a = \text{diag}(\mathbf{h}_a)\mathbf{s}+\mathbf{n}_a,\ \ \ a=1,\cdots, R, \IEEEeqnarraynumspace
\end{IEEEeqnarray}
where $[\mathbf{h}_a]_i\sim \mathcal{CN}(\mathbf{0},1)$ is the channel coefficient of the $i$th symbol, $[\mathbf{s}]_i$, and $\mathbf{n}_a\sim \mathcal{CN}(\mathbf{0},N_0\mathbf{I}_V)$ is the vector of additive white Gaussian noise (AWGN) samples. Note that, in writing \eqref{Eq_received_sig}, it is assumed that the fading coefficient is constant (channel fades slowly) over one symbol duration. However, the duration of the vector of symbols, $\mathbf{s}$, can be much larger than the channel coherence time. Due to the presence of the interleaver/deinterleaver in the system, the fading coefficients can be modeled as independent and
identically distributed (i.i.d.) Rayleigh random variables \cite{Caire98}. Also, it is assumed that channels across the receive antennas are i.i.d.

\begin{example}
	For a wireless terminal operating at a speed of $v$ km/hr and a frequency of $f_{\rm o}$ GHz, the maximum Doppler frequency is $f_{D}\approx \frac{vf_{\rm o}}{1.08}$ Hz. This leads to a coarse estimate $T_{\rm coh}\approx\frac{1}{4f_{D}}=\frac{1}{4vf_{\rm o}}$ of the channel coherence time [41]. Bits separated in time by more than one $T_{\rm coh}$ are almost independently affected by channel fading. Consider a single-carrier system with bandwidth $B$ Hz. The system modulates a group of $m$ coded bits over a signal of duration $B^{-1}$ (seconds), chosen from a constellation of size $2^m$. As such, the approximate number of bits affected by one coherence time is $N_{\rm coh}=mBT_{\rm coh}\approx \frac{mB}{4vf_{\rm o}}$. Consider two codewords of length $L_c$ and Hamming distance $d$, where $L_c>>d$, and $L_c>>\frac{mB}{4vf_{\rm o}}$. Under these assumptions, a random interleaver of length $L_c$ will separate (with high probability) the location of the $d$ bits responsible for the Hamming distance, well enough, such that those bits are affected by independent channel coefficients.
\end{example}

The receiver employs a suboptimal, iterative processing method that is based on individually optimal, but separate demodulation and channel decoding blocks. The presence of the interleaver and deinterleaver in the iterative receiver is quite obvious and necessary to properly reorder the exchanged information between the EB SISO demodulator and SISO decoder. The operations of the these two SISO blocks are described in the following.

\subsubsection{EB SISO Demodulator} The inputs to the EB SISO demodulator are the received signals from $R$ antennas, and the \emph{a priori} information about the vector of interleaved coded bits $\mathbf{v}$ that is provided by the SISO channel decoder in the previous iteration, denoted as $\mathbf{d}_b$ for $b\in\{0,1\}$. The SISO demodulator produces two vectors $\mathbf{P}_b$, where $b\in\{0,1\}$, of extrinsic \emph{a posteriori} information about $\mathbf{v}$. Specifically, consider the $k$th bit in $\mathbf{v}$, where $k=1,\cdots, L_c$. Write $k=[v(k)-1]m + w(k)$, where $1\leq v(k) \leq V$ and $1\leq w(k)\leq m$. In other words, $v(k)=(k-1)\setminus m+1$ and $w(k)=\text{rem}(k-1,m)+1$. This means that the $k$th bit in $\mathbf{v}$ is carried as the $w(k)$th labeling bit of the $v(k)$th symbol in $\mathbf{s}$. The value in $\mathbf{P}_b$ that corresponds to the $k$th bit in $\mathbf{v}$ is then calculated as
\begin{IEEEeqnarray}{c}
	\label{Eq_SISO_extrinsic}
	[\mathbf{P}_b]_k  =  \frac{{\rm Pr}\left([\mathbf{v}]_k=b\ | \  [\mathbf{y}_1]_{v(k)},\cdots,[\mathbf{y}_R]_{v(k)}\right)}{[\mathbf{d}_b]_k}, \IEEEeqnarraynumspace
\end{IEEEeqnarray}
where $b\in\{0,1\}$. The numerator of \eqref{Eq_SISO_extrinsic} can be further computed as
\begin{IEEEeqnarray}{lll}
	\label{Eq_SISO_demod_num}
	{\rm Pr}([\mathbf{v}]_k=b | [\mathbf{y}_1]_{v(k)},\cdots,[\mathbf{y}_R]_{v(k)}) \IEEEnonumber\\
	\propto  \sum_{s\in \Psi_b^{w(k)}} f([\mathbf{y}_1]_{v(k)},\cdots,[\mathbf{y}_R]_{v(k)} | [\mathbf{s}]_{v(k)}=s)
 \times {\rm Pr}([\mathbf{s}]_{v(k)}=s),
	\IEEEeqnarraynumspace
\end{IEEEeqnarray}
where $\Psi_b^{w(k)}$ represents the set of $\frac{M+1}{2}$ constellation points having the labeling bit at position $w(k)$ equal to $b$, and the conditional probability density function (pdf) $f(\cdot | \cdot)$ is defined as
\begin{IEEEeqnarray}{ccc}
	\label{Eq_SISO_ML}
	f([\mathbf{y}_1]_v,\cdots,[\mathbf{y}_R]_v \ |\  [\mathbf{s}]_v=s)= \frac{\exp\left(-\frac{\sum_{a=1}^{R}\left| [\mathbf{y}_a]_v \right|^2}{s^2+N_0}\right)}{\pi^R(s^2+N_0)^R}. \IEEEeqnarraynumspace
\end{IEEEeqnarray}
Note that, due to the unavailability of CSI, the effect of signal points on the likelihood function appears in the variance.
	

At the start, no \emph{a priori} information of the encoded bits is given by the channel decoder. Hence, the receiver is open loop (the feedback link in Fig. \ref{fig_block_BICM_ID} is ignored). As such, the symbols in $\mathbf{s}$ are equally likely, i.e., the \emph{a prior} distribution is ${\rm Pr}([\mathbf{s}]_v=s)=\frac{1}{M+1}$ for all $v$ and $s$. It follows from \eqref{Eq_SISO_extrinsic}, \eqref{Eq_SISO_demod_num}, and \eqref{Eq_SISO_ML} that the output of the SISO demodulator before iterations begin is given by
 \begin{IEEEeqnarray}{c}
 	\label{Eq_SISO_first}
 	[\mathbf{P}_b]_k\propto \sum_{s\in \Psi_b^{w(k)}} \frac{\exp\left(-\frac{\sum_{a=1}^{R}\left| [\mathbf{y}_a]_{v(k)} \right|^2}{s^2+N_0}\right)}{(s^2+N_0)^R}. \IEEEeqnarraynumspace
 \end{IEEEeqnarray}
Starting from the first iteration, the SISO decoder provides the demodulator with \emph{a priori} information of the encoded bits in $\mathbf{v}$, which is stored in vectors $\mathbf{d}_b$, $b\in \{0,1\}$. Noting that the $m$ bits in $\mathbf{v}$ that are mapped to the $v(k)$th symbol in $\mathbf{s}$ are at positions $[v(k)-1] m + j$, $j=1,\cdots, m$, \eqref{Eq_SISO_extrinsic} can be computed as
\begin{IEEEeqnarray}{ccc}
	\label{Eq_SISO_extrinsic_next_iter}
[\mathbf{P}_b]_k   \propto
 \sum_{s\in \Psi_b^{w(k)}} \frac{\exp\left(-\frac{\sum_{a=1}^{R}\left| [\mathbf{y}_a]_{v(k)} \right|^2}{s^2+N_0}\right)}{(s^2+N_0)^R} \times \prod_{{j=1}\atop{j\neq w(k)}}^m [\mathbf{d}_{b_j(s)}]_{j+k-w(k)},\IEEEeqnarraynumspace
\end{IEEEeqnarray}
where  $b_j(s)\in \{0,1\}$ is the value of the $j$th labeling bit of symbol $s$. Note that, in obtaining \eqref{Eq_SISO_extrinsic_next_iter} we assume that the bits carried by each symbol $s$ are independent, and hence the probability ${\rm Pr}([\mathbf{s}]_{v(k)}=s)$ can be computed as the product of probabilities of the labeling bits of $s$. This is a reasonable assumption thanks to the random interleaver.

\subsubsection{SISO Decoder} The function of the decoder is to update the deinterleaved extrinsic information $\mathbf{q}_b$, $b\in \{0,1\}$, and feed it back to the demodulator for the next iteration. In this regard, the SISO module utilizes the trellis structure of the $(n_c,k_c,\nu_c)$ convolutional code $\textbf{c}$ to compute the maximum a posteriori probability (MAP) estimates $[\mathbf{b}_b]_k=\Pr\left(\mathbf{q}_0, \mathbf{q}_1 \ |\ [\textbf{c}]_k=b \right)$. 
The trellis diagram, defined over $\breve{K}=L_c/n_c$ time instants, plays an important role, as it facilitates a very efficient recursive approach to computing $\Pr\left(\mathbf{q}_0, \mathbf{q}_1 \ |\ [\textbf{c}]_k=b \right)$. Let us denote a trellis edge by $e$, and its starting and ending states by $S^{\rm S}(e)$ and $S^{\rm E}(e)$, respectively. A sequence $\{c^j_{\breve{k}}(e)\}$ of coded bits of length $n_c$ is associated with edge $e$ at time instant $\breve{k}$. That is, $c_{\breve{k}}^i(e)\in \{0,1\}$ is the $i$th output of the convolutional encoder at time $\breve{k}$, corresponding to the state transition $S^{\rm S}(e)$ to $S^{\rm E}(e)$ ($\breve{k}=1,\cdots,K$ and $i=1,\cdots,n_c$). Then, it is simple to show that \cite{Benedetto97,Book-Shulin}
\begin{IEEEeqnarray}{ccc}
	\label{Eq_SISO_decoder}
	[\mathbf{b}_b]_{k(j)} \propto H_c \sum_{e: c_{\breve{k}}^j(e)=b} A_{\breve{k}-1} (S^{\rm S}(e)) \cdot \left(\prod\limits_{\substack{i=1 \\ i\neq j}}^{n_c}[\mathbf{q}_{c^i_{\breve{k}}(e)}]_{k(i)}\right) \cdot B_{\breve{k}}(S^{\rm E}(e)),
	\IEEEeqnarraynumspace
\end{IEEEeqnarray}
where $k(l)=(\breve{k}-1)n_c+l$, for every $l\in \{1,\cdots,n_c\}$, and $H_c$ is the normalization constant, making $[\mathbf{b}_0]_{k(j)}=1-[\mathbf{b}_1]_{k(j)}$. Note that in \eqref{Eq_SISO_decoder}, $\mathbf{q}_{c^i_{\breve{k}}(e)}$ is equal to $\mathbf{q}_0$  (or $\mathbf{q}_1$) if $c^i_{\breve{k}}(e)=0$ (or $c^i_{\breve{k}}(e)=1$). Furthermore, the quantities $A_{\breve{k}}(\cdot)$ and $B_{\breve{k}}(\cdot)$ can be obtained through the forward and backward recursion, respectively, as
\begin{IEEEeqnarray}{ccc}
	\label{Eq_forw_rec}
	A_{\breve{k}}(s)=\!\!\!\! \sum_{e:S^{\rm E}(e)=s}A_{\breve{k}-1}(S^{\rm S}(e))\prod\limits_{i=1}^{n_c}[\mathbf{q}_{c^i_{\breve{k}}(e)}]_{k(i)},\ \ B_{\breve{k}}(s)=\!\!\!\! \sum_{e:S^{\rm S}(e)=s}B_{\breve{k}+1}(S^{\rm E}(e))\prod\limits_{i=1}^{n_c}[\mathbf{q}_{c^i_{\breve{k}}(e)}]_{k(i)+n_c},
	\IEEEeqnarraynumspace
\end{IEEEeqnarray}
with initial values $A_0(s)=B_{\breve{K}}(s)=1$ if $s=S_0$, and $A_0(s)=B_K(s)=0$, otherwise ($S_0$ is the all-zero state that the trellis diverges from and converges to).


When the updated extrinsic information $\mathbf{b}_b$ passes through the interleaver, it becomes the \emph{a priori} information on the interleaved coded bits in $\mathbf{v}$. 
Note that after any desired number of iterations, the final decision $\hat{\mathbf{u}}$ on the information vector $\mathbf{u}$ is made by making hard decisions on the elements of  $\mathbf{q}_b$.

\begin{remark}
	When the decision on $\mathbf{u}$ is made without performing any iterations, the system is called \emph{feedback-free} BICEM, or simply BICEM. As we will discuss later, the performance of the BICEM system in the FF mode has a significant impact on the performance of the BICEM-ID system.
\end{remark}

\begin{remark}
	(System complexity)  The complexity of the soft output demodulator is insignificant compared to the SISO decoder, which needs forward and backward recursions. So, the BICEM-ID's receiver complexity is dominated by the SISO decoder, which has the same structure as the one employed in standard BICM-ID schemes \cite{Chindapol01}. The number of visited edges of the code trellis per decoded bit is a good measure of the decoder complexity \cite{Chindapol01}. This is roughly three times the Viterbi algorithm's complexity of $\frac{2^{k_c+\nu_c}}{k_c}$ per trellis section \cite{Book-Shulin,Chindapol01}. In conclusion, the code's constraint length $\nu_c$ determines the order of complexity. Since the number of arithmetic operations of the decoding algorithm is the same at each trellis section, the decoder's complexity grows \emph{linearly} by the code length $L_c$. This is in contrast to the complexity of the optimal receiver of Fig. \ref{fig_general_diag}, whose complexity increases \emph{exponentially} by the length $L$ of the input vector $\textbf{u}$. 
\end{remark}
\begin{remark}
	Instead of convolutional codes, one can employ block codes such as low-density parity-check (LDPC) codes. In such cases, the efficient soft decoding is feasible by recursive processing over the Tanner graph of the code. As the main objective of this paper is to investigate turbo processing of information when EB modulator and non-coherent demodulator are employed, the comparative study of BICEM-ID under different encoding/decoding techniques, also incorporating non-coherent EBM into state-of-the-art schemes such as protograph LDPC BICM and spatially-coupled LDPC BICM \cite{Yang21iot,Fang19TCOM,Yang20TVT}, are interesting topic beyond the scope of this paper.
\end{remark}

\section{Upper-Bounds on the Error Performance and Design Guidelines}\label{sec-ana}

Since the non-coherent BICEM-ID system is developed based on the same principle of the conventional coherent BICM-ID system, they also share similar principles in performance analysis. Specifically, there are two metrics that determine the asymptotic performance of the BICEM-ID system. The first is the asymptotic bit error rate (BER) of the FF BICEM. This metric, which is dependent on the signal mapping, evaluates how well the system works before iterations start. The second metric is a bound on the \emph{error-free feedback} performance of BICEM-ID. As the name suggests, the EFF performance is based on the assumption that the first iteration works well and the feedback from the decoder to the demodulator with subsequent iterations eventually becomes error-free \cite{Chindapol01,Li02}. This performance is essentially the ultimate performance to which BICEM-ID converges at high SNRs. As with BICM-ID, it will be seen that signal mapping from the coded bits in $\mathbf{v}$ to the energy levels in $\mathbf{s}$ also plays a crucial role in determining the error performance of BICEM-ID.

A general expression for the union bound on the BER of BICEM-ID can be written as \cite{Caire98}
 \begin{IEEEeqnarray}{c}
	\label{Eq_General_union}
	P_{\rm UB}=\frac{1}{k_{\rm c}}\sum_{d=d_{\rm H}}^{\infty}c_dg(d,\Psi,\xi), \IEEEeqnarraynumspace
\end{IEEEeqnarray}
where $c_d$ is the total information weight of all error events at Hamming distance $d$ and $d_{\rm H}$ is the free Hamming distance of the code. Also, $g(d,\Psi,\xi)$ is the average PEP, which is a function of the constellation $\Psi$, Hamming distance $d$, and the mapping rule $\xi(\cdot)$. 

Let $\mathbf{c}$ and $\hat{\mathbf{c}}$ denote transmitted and detected codewords, respectively, with a Hamming distance $d$ between them. These codewords correspond to symbol vectors $\mathbf{s}$ and $\hat{\mathbf{s}}$, respectively. The function $g(d,\Psi,\xi)$ can be obtained by taking the average of the PEP $\Pr(\mathbf{s}\rightarrow \hat{\mathbf{s}}|\mathbf{s})$ over all valid pairs $(\mathbf{s},\hat{\mathbf{s}})$ with respect to the mapping $\xi$. Without loss of generality, suppose $\mathbf{c}$ and $\hat{\mathbf{c}}$ differ in the first $d$ bits. Thanks to the use of a sufficiently long interleaver, we can assume that each of these $d$ bits in $\mathbf{c}$ (and $\hat{\mathbf{c}}$) is modulated by (mapped to) a different symbol in $\mathbf{s}$ (and $\hat{\mathbf{s}}$). For instance, with $L_c=9000$ ($L_c=12000$), and $m=3$ ($m=2$), the probability that $d=10$ bits are modulated by $10$ different symbols is approximately $0.99$ ($0.996$). Hence, without affecting analytical results, in the rest of this section we can consider that $\mathbf{s}$ and $\hat{\mathbf{s}}$ include only $d$ symbols, and the label of $[\mathbf{s}]_v$ is different from that of $[\hat{\mathbf{s}}]_v$ ($v=1,\cdots,d$) in only one bit.

The PEP $\Pr(\mathbf{s}\rightarrow \hat{\mathbf{s}}|\mathbf{s})$ can be computed as follows:
 \begin{IEEEeqnarray}{cc}
	\label{Eq_PEP_expression}
	\Pr(\mathbf{s}\rightarrow \hat{\mathbf{s}}|\mathbf{s})= \Pr\{\ln\left(f(\mathbf{y}_1,\cdots,\mathbf{y}_R | \mathbf{s})\right) < \ln\left(f(\mathbf{y}_1,\cdots,\mathbf{y}_R | \hat{\mathbf{s}})\right) |
	 \ \mathbf{s}\}, \IEEEeqnarraynumspace
\end{IEEEeqnarray}
where, for $\mathbf{s}'\in \{\mathbf{s},\hat{\mathbf{s}}\}$
\begin{IEEEeqnarray}{cc}
	\label{Eq_Likelihood_func}
	f(\mathbf{y}_1,\cdots,\mathbf{y}_R | \mathbf{s}')=
	 \frac{\exp\left(-\sum_{a=1}^{R}\sum_{v=1}^{d}\frac{|[\mathbf{y}_a]_v |^2}{[\mathbf{s}']_v^2+N_0}\right)}{\prod_{v=1}^{d}\left([\mathbf{s}']_v^2+N_0\right)^R}.
	 \IEEEeqnarraynumspace
\end{IEEEeqnarray}
Now, consider the $v$th element of $\mathbf{s}'$, i.e., $[\mathbf{s}']_v$. According to \eqref{Eq_singleshot_levels}, we have $[\mathbf{s}']^2_v+N_0=N_0r^{q'_{v}}$ for some $q'_{v}\in \{0,\cdots,M\}$. This means that we can equivalently represent $[\mathbf{s}']_v$ by $q'_{v}$. Substituting this equality into \eqref{Eq_Likelihood_func} and taking logarithm yield
\begin{IEEEeqnarray}{cc}
	\label{Eq_Likelihood_func_subs}
	\ln\left(f(\mathbf{y}_1,\cdots,\mathbf{y}_R |  \mathbf{s}')\right)\propto
	-R\sum_{v=1}^{d} q'_{v}\ln(r)-\sum_{a=1}^{R}\sum_{v=1}^{d}\frac{|[\mathbf{y}_a]_v |^2}{N_0r^{q'_{v}}}.
	\IEEEeqnarraynumspace
\end{IEEEeqnarray}
Substituting \eqref{Eq_Likelihood_func_subs} into \eqref{Eq_PEP_expression} results in
\begin{IEEEeqnarray}{cc}
	\label{Eq_PEP_expression_subs}
	\hspace{-0.5cm}\Pr(\mathbf{s}\rightarrow \hat{\mathbf{s}}|\mathbf{s})= \Pr\left\{\sum_{a=1}^{R}\sum_{v=1}^{d}\left(\frac{1}{N_0r^{q_{v}}}-\frac{1}{N_0r^{\hat{q}_{v}}}\right)|[\mathbf{y}_a]_v |^2 >\right.  \left. R\ln(r)\sum_{v=1}^{d}\left(\hat{q}_{v}-{q}_{v}\right)\; \right\}, \IEEEeqnarraynumspace
\end{IEEEeqnarray}
where we have substituted $q'_{v}=q_{v}$ for $\mathbf{s}'=\mathbf{s}$, and  $q'_{v}=\hat{q}_{v}$ for $\mathbf{s}'=\hat{\mathbf{s}}$.

Note that in \eqref{Eq_PEP_expression_subs},
\begin{IEEEeqnarray}{cc}
	\label{Eq_y_ki}
	[\mathbf{y}_a]_v \sim {\mathcal C}{\mathcal N}\left(\mathbf{0},[\mathbf{s}]^2_v+N_0\right)= {\mathcal C}{\mathcal N}\left(\mathbf{0},N_0r^{q_v}\right), \IEEEeqnarraynumspace
\end{IEEEeqnarray}
which is equivalent to the distribution of the random variable $\sqrt{N_0r^{q_v}}Z_{v,a}$, with $Z_{v,a}\sim {\mathcal C}{\mathcal N}\left(\mathbf{0},1\right)$.
Consequently, \eqref{Eq_PEP_expression_subs} can be written in a simplified form as
 \begin{IEEEeqnarray}{cc}
	\label{Eq_PEP_expression2}
		\Pr(\mathbf{s}\rightarrow \hat{\mathbf{s}}|\mathbf{s})=	 \Pr\left\{\sum_{a=1}^{R}\sum_{v=1}^{d}\right.\left(1-r^{q_{v}-\hat{q}_{v}}\right)| Z_{v,a} |^2
		  \left.> R\ln(r)\sum_{v=1}^{d}\left(\hat{q}_{v}-q_{v}\right)\right\}. \IEEEeqnarraynumspace
\end{IEEEeqnarray}
The computation of PEP in traditional (coherent) communication, where the CSI is available at the receiver, is usually straightforward. Unfortunately, exact solution to \eqref{Eq_PEP_expression2} is too complicated, if not impossible. Hence, our strategy changes to finding tight upperbounds on the PEP. By applying the Chernoff inequality to \eqref{Eq_PEP_expression2}, it is shown in Appendix \ref{appxA} that
 \begin{IEEEeqnarray}{ll}
	\label{Eq_PEP_Chernoffbound}
	\Pr(\mathbf{s}\rightarrow \hat{\mathbf{s}}|\mathbf{s}) &\leq
	\frac{\prod_{a=
			1}^{R}\prod_{v=1}^{d} {\rm E}_{Z_{v,a}}[\exp\left(t\left(1-r^{q_{v}-\hat{q}_{v}}\right)| Z_{v,a} |^2\right)]}{\prod_{v=1}^{d}\exp\left(tR\ln(r)\left(\hat{q}_{v}-q_{v}\right)\right)}\IEEEnonumber \\
&= \prod_{v=1}^{d}\left(\frac{r^{t(q_v-\hat{q}_v)}}{1-t(1-r^{q_v-\hat{q}_v})}\right)^R,\ \ \ \ 0\leq t\leq \frac{1}{1-r^{-M}}, \IEEEeqnarraynumspace
\end{IEEEeqnarray}
where ${\rm E}_X[\cdot]$ means the expectation with respect to random variable $X$, and $t$ is the range over which the expectation is well-defined.

Next, the average PEP, i.e., the function $g(d,\Psi,\xi)$ in \eqref{Eq_General_union}, can be obtained by averaging \eqref{Eq_PEP_Chernoffbound} over all \emph{relevant} realizations of $(q_{v}, \hat{q}_{v})$. In the following, we perform such averaging in two special cases: FF and EFF. To this end, we shall introduce the following definitions.

Let $s^{[l]}=(r^l-1)N_0$, $l=0,\cdots,M$, denote the $(l+1)$th signal point in $\Psi$. Furthermore, let $b$ be the $w$th labeling bit of $s^{[l]}$, where $b\in\{0,1\}$ and $w=1,\cdots,m$. With respect to $s^{[l]}$, we define $s^{[\kappa(l,w)]}$ as the signal point in $\Psi$ having the following properties: (i) It belongs to $\Psi_{\bar{b}}^w$, i.e., its $w$th labeling bit is $\bar{b}$, the complement of $b$, and (ii) Among all signal points in $\Psi_{\bar{b}}^w$, it is the one that is nearest to $s^{[l]}$ in terms of their indices, $|\kappa(l,w) - l|$. On the other hand, $s^{[\varrho(l,w)]}$ is defined as the signal point in $\Psi$ whose label bits are the same as those of $s^{[l]}$, except at position $w$. In the following, we give an example to show the values of $\kappa(l,w)$ and $\varrho(l,w)$. For convenience, we represent the mapping rule $s=\xi(\cdot)$ by a column vector $\mathbf{L}$ of length $M+1$, where $[\mathbf{L}]_{l+1}$ is the decimal representation of the binary label of $s^{[l]}$, $l=0,\cdots,M$, in which the rightmost bit in the label is the most significant bit. For instance $[\mathbf{L}]_{l+1}=4$ is the decimal representation of the label $``0\ 0\ 1"$ (the bit farthest to right has the largest value).

\begin{figure*}[h]
	\centering
	\includegraphics[width=5.25in]{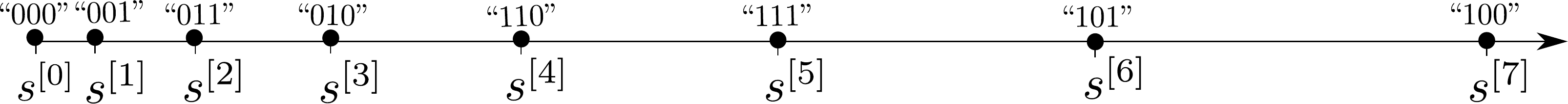}
	\caption{Signal points and Gray labeling considered in Example \ref{Ex1}. Note that $s^{[l]}=\sqrt{p_l}=\sqrt{(r^l-1)N_0}$}
	\label{fig_example}
\end{figure*}

\begin{example} \label{Ex1}
Consider a $8$-ary constellation $\Psi=\{s^{[0]},\cdots,s^{[7]}\}$, where $s^{[l]}$, $l=0,\cdots,7$, is given by \eqref{Eq_singleshot_levels}. Figure \ref{fig_example} shows the signal points when a Gray mapping is applied. The mapping in Fig. \ref{fig_example} can be represented by $\mathbf{L}_{\rm Gray}=[0\ 4\ 6\ 2\ 3\ 7\ 5\ 1]^\top$. Note that, according to \eqref{Eq_singleshot_levels}, $\Psi$ is a \emph{geometric sequence} \cite{Gao2020} in which the Euclidean distance between adjacent signal points increases with $l$. Table \ref{table_EX1} gives the values of $\kappa(l,w)$ and $\varrho(l,w)$ for all combinations of $l$ and $w$. For example, the entries in the table for $(l,w)=(4,2)$ are $\kappa(l,w)=6$ and $\varrho(l,w)=7$.

It is important to emphasize that $s^{[\kappa(l,w)]}$ is the point in $\Psi_{\bar{b}}^w$ that is nearest to $s^{[l]}$ in terms of index distance. This is in contrast to traditional BICM-ID schemes, where nearest neighbor is defined by Euclidean metric. To illustrate this point, consider signal $s^{[4]}$ and its second labeling bit ($w=2$), which is $b=1$. For all signal points in $\Psi_0^2=\{s^{[0]},s^{[1]},s^{[6]},s^{[7]}\}$, $s^{[6]}$ is nearest to $s^{[4]}$ because $\argmin_{\hat{l}\in\{0,1,6,7\}} | 4-\hat{l}|=6$. Yet, $s^{[6]}$ is not the signal in $\Psi_0^2$ that is nearest to $s^{[4]}$ in the Euclidean distance. For example, at $\gamma=\frac{1}{N_0}=20$dB, the solution to \eqref{Eq_all_one_poly} results in $r=2.41$. Then, from \eqref{Eq_singleshot_levels}, $|~s^{[4]}-s^{[1]}~| = |~\sqrt{(r^4-1)N_0}-\sqrt{(r-1)N_0}~| \approx 0.45$ which is smaller than $| s^{[4]}-s^{[6]} |=| \sqrt{(r^4-1)N_0}-\sqrt{(r^6-1)N_0} | \approx 0.82$. Hence, among all the signal points in $\Psi_0^2$, $s^{[1]}$ is the nearest neighbor to $s^{[4]}$ in the Euclidean distance.  $\blacksquare$
\end{example}

With the above definitions of $\kappa(l,w)$ and $\varrho(l,w)$, averaging \eqref{Eq_General_union} in the two cases of FF and EFF can be performed as follows. In the FF scenario, bit $w$ of the transmitted symbol $s^{[q_v]}\in \Psi_b^w$ is wrongly detected if the demodulated symbol is in $\Psi_{\bar{b}}^w$. Appendix \ref{appxB} shows that, in analyzing the error performance of the proposed BICEM system, the ``nearest neighbor'' should be defined in terms of symbol indices. Using this fact, and by employing the nearest-neighbor bounding technique, we can average the PEP in \eqref{Eq_General_union} over all ``nearest-neighbor'' pairs $(s^{[q_v]},s^{[\kappa(q_v,w)]})$, or $(q_v,\kappa(q_v,w))$. On the other hand, in the EFF scenario, the feedback from the channel decoder to the demodulator provides a complete knowledge of all bits in the label of $s^{[q_v]}$, except bit $w$. Hence, the average PEP is obtained by averaging \eqref{Eq_PEP_Chernoffbound} over all pairs $(q_v,\varrho(q_v,w))$, $q_v=0,\cdots,M$, and $w=1,\cdots,m$.

\begin{table}[tb]
	\centering
	\caption{Functions $\kappa(l,w)$ and $\varrho(l,w)$ for the Gray-labeled $8$-ary constellation in Example $1$.}
	\label{table_EX1}
	\begin{tabular}{|c|cccc|cccc|c}
		\cline{1-9}
		\backslashbox{$w$}{$l$} & \multicolumn{1}{c|}{$0$} & \multicolumn{1}{c|}{$1$} & \multicolumn{1}{c|}{$2$} & $3$ & \multicolumn{1}{c|}{$4$} & \multicolumn{1}{c|}{$5$} & \multicolumn{1}{c|}{$6$} & $7$ & \multicolumn{1}{l}{}                      \\ \hline
		1   & \multicolumn{4}{c|}{$4$}                                                        & \multicolumn{4}{c|}{$3$}                                                        & \multicolumn{1}{c|}{\multirow{3}{*}{$\kappa(l,w)$}} \\ \cline{1-9}
		2   & \multicolumn{2}{c|}{$2$}                           & \multicolumn{2}{c|}{$1$}      & \multicolumn{2}{c|}{$6$}                           & \multicolumn{2}{c|}{$5$}      & \multicolumn{1}{c|}{}                     \\ \cline{1-9}
		3   & \multicolumn{1}{c|}{$1$}  & \multicolumn{1}{c|}{$0$}  & \multicolumn{1}{c|}{$3$}  & $2$  & \multicolumn{1}{c|}{$5$}  & \multicolumn{1}{c|}{$4$}  & \multicolumn{1}{c|}{$7$}  & $6$  & \multicolumn{1}{c|}{}                    \\ \hline \hline
		1   & \multicolumn{1}{c|}{$7$}  & \multicolumn{1}{c|}{$6$}  & \multicolumn{1}{c|}{$5$}  & $4$  & \multicolumn{1}{c|}{$3$}  & \multicolumn{1}{c|}{$2$}  & \multicolumn{1}{c|}{$1$}  & $0$  & \multicolumn{1}{c|}{\multirow{3}{*}{$\varrho(l,w)$}} \\ \cline{1-9}
		2   & \multicolumn{1}{c|}{$3$}  & \multicolumn{1}{c|}{$2$}  & \multicolumn{1}{c|}{$1$}  &  $0$ & \multicolumn{1}{c|}{$7$}  & \multicolumn{1}{c|}{$6$}  & \multicolumn{1}{c|}{$5$}  & $4$  & \multicolumn{1}{c|}{}                     \\ \cline{1-9}
		3   & \multicolumn{1}{c|}{$1$}  & \multicolumn{1}{c|}{$0$}  & \multicolumn{1}{c|}{$3$}  & $2$  & \multicolumn{1}{c|}{$5$}  & \multicolumn{1}{c|}{$4$}  & \multicolumn{1}{c|}{$7$}  & $6$  & \multicolumn{1}{c|}{}                     \\ \hline
	\end{tabular}
\end{table}

From \eqref{Eq_PEP_Chernoffbound}, a unified expression for the average PEP is
\begin{IEEEeqnarray}{ll}
	\label{Eq_average_PEP}
	g_{\Phi}(d,\Psi,\xi)&\leq \prod_{v=1}^{d}{\rm E}_{q_{v},\hat{q}_v}\left[\left(\frac{r^{t(q_v-\hat{q}_v)}}{1-t(1-r^{q_v-\hat{q}_v})}\right)^R\right]
=\left({\rm E}_{q_{v},\hat{q}_v}\left[\left(\frac{r^{t(q_v-\hat{q}_v)}}{1-t(1-r^{q_v-\hat{q}_v})}\right)^R\right]\right)^d\IEEEnonumber\\
&	= \left(\frac{1}{m(M+1)}\sum_{l=0}^{M}\sum_{w=1}^{m}\left(\frac{r^{t(l-\Phi(l,w))}}{1-t(1-r^{l-\Phi(l,w)})}\right)^R\right)^d, \IEEEeqnarraynumspace
\end{IEEEeqnarray}
where ${\rm E}_{q_{v},\hat{q}_v}(\cdot)$ takes the average over all relevant pairs $(q_{v},\hat{q}_v)$, whereas the symbol $\Phi$ (in both the subscript of function $g(\cdot)$ and the function $\Phi(l,w)$ itself) is replaced with $\kappa$ or $\varrho$ for the FF and EFF cases, respectively.

It is shown in Appendix \ref{appxC} that for the EFF case, \eqref{Eq_average_PEP} gives \emph{the tightest upper-bound} on $g_{\varrho}(d,\Psi,\xi)$ when $t=\frac{1}{2}$. On the other hand, the value of $t$ that maximizes the upper-bound in \eqref{Eq_average_PEP} in the FF case depends on the signal mapping. However, numerical results show that the upper-bound of $g_{\kappa}(d,\Psi,\xi)$ in \eqref{Eq_average_PEP} has its minimum at $t$ equal to or very close to $\frac{1}{2}$. Therefore, substituting $t=\frac{1}{2}$ into \eqref{Eq_average_PEP} results in
\begin{IEEEeqnarray}{cc}
	\label{Eq_average_PEP_tightset}
g_{\Phi}(d,\Psi,\xi)
	\leq \delta^d_{\Phi}(\Psi,\xi),
	\IEEEeqnarraynumspace
\end{IEEEeqnarray}
{where
\begin{IEEEeqnarray}{lll}
	\label{Eq_delta}
	\delta_{\Phi}(\Psi,\xi)&
	=&\frac{1}{m(M+1)}\sum_{l=0}^{M}\sum_{w=1}^{m}\left(\frac{2}{r^{-\frac{1}{2}| l-\Phi(l,w)|}+r^{\frac{1}{2}| l-\Phi(l,w)|}}\right)^R\IEEEnonumber\\
	&=&\frac{1}{m(M+1)}\sum_{l=0}^{M}\sum_{w=1}^{m}{\rm cosh}^{-R}\left(\frac{| l-\Phi(l,w)|\ln(r)}{2}\right).
	\IEEEeqnarraynumspace
\end{IEEEeqnarray}
To write \eqref{Eq_delta} in a simpler form, let us use the fact that $1 \leq|l-\Phi(l,w)|\leq M$, $l\in \{0,\cdots,M\}$, $w\in \{1,\cdots,m\}$. Then, \eqref{Eq_delta} can be reformulated as
\begin{IEEEeqnarray}{cc}
	\label{Eq_delta_new}
	\delta_{\Phi}(\Psi,\xi)=	\frac{1}{m(M+1)}
	\sum_{j=1}^{M}N_j\cdot {{\rm cosh}^{-R}\left(\frac{j\ln(r)}{2}\right)},
	\IEEEeqnarraynumspace
\end{IEEEeqnarray}
where $N_j$ is the number of pairs $(l,w)$ for which $|l-\Phi(l,w)|=j$. Suppose that $N_j\neq 0$ (i.e., $N_j$ is a positive integer) at $K$ indices $j=n_1, n_2,\cdots,n_K$, such that $1\leq n_1< n_2< \cdots < n_K\leq M$. Then, $n_k$, $k=1,\cdots,K$ identifies the $n_k$th nearest neighbor, and $N_{n_k}$ is the number of the $n_k$th nearest neighbors within the constellation. It follows that \eqref{Eq_delta_new} can be written as
\begin{IEEEeqnarray}{cc}
	\label{Eq_delta_new2}
	\delta_{\Phi}(\Psi,\xi)=	\frac{1}{m(M+1)}
	\sum_{k=1}^{K}N_{n_k}\cdot {{\rm cosh}^{-R}\left(\frac{n_k\ln(r)}{2}\right)}.
	\IEEEeqnarraynumspace
\end{IEEEeqnarray}
To further simplify \eqref{Eq_delta_new2}, the following lemma is useful.} \begin{lemma}\label{lemma1}
The value $r$, which determines the optimal power levels, is approximated as $r\approx (\varpi \gamma_{\rm b})^{\frac{1}{M}}$ at large SNRs, where $\varpi=\frac{mk_{\rm c}}{n_{\rm c}}(M+1)$, and $\gamma_{\rm b}$ is the \emph{SNR per bit}.

{\prf Note that the solution to \eqref{Eq_all_one_poly} is equivalent to the solution to $\frac{r^{M+1}-1}{r-1}=(M+1)(\gamma+1)$. In the limit of large SNRs, we have $r\gg 1$. In other words, $\frac{r^{M+1}-1}{r-1}\rightarrow r^M$, when $\gamma\rightarrow \infty$. This results in $r\approx ((M+1)\gamma)^{\frac{1}{M}}$. Since each constellation point carries $\frac{k_{\rm c}}{n_{\rm c}}m$ information bits, we can write $\gamma=\frac{mk_{\rm c}}{n_{\rm c}}\gamma_{\rm b}$, where $\gamma_{\rm b}$ is the SNR per information bit. Consequently, $r\approx \left(\frac{mk_{\rm c}(M+1)}{n_{\rm c}}\gamma_{\rm b}\right)^{\frac{1}{M}}=\left(\varpi \gamma_{\rm b}\right)^{\frac{1}{M}}$. $\blacksquare$}
\end{lemma}

{Under the large SNR regime, we can apply the above lemma to \eqref{Eq_delta_new2} to arrive at
\begin{IEEEeqnarray}{lll}
	\label{Eq_delta_new3}
	\delta_{\Phi}(\Psi,\xi)&\overset{{\rm (i)}}{\approx}&	\frac{1}{m(M+1)}
	\sum_{k=1}^{K}N_{n_k} {{\rm cosh}^{-R}\left(\frac{n_k\ln(\varpi\gamma_{\rm b})}{2M}\right)}\hfill\IEEEnonumber\\
	&\overset{{\rm (ii)}}{\approx}& \frac{1}{m(M+1)}\sum_{k=1}^{K}N_{n_k} {{\rm cosh}^{-R}}{\left(\frac{n_k([\varpi]_{\rm dB}+[\gamma_{\rm b}]_{\rm dB})}{8.7 M}\right)}\hfill\IEEEnonumber\\
	&=&\frac{N_{n_1}\Xi(\gamma_{\rm b},M)}{m(M+1)} {\rm cosh}^{-R}\left(\frac{n_1([\varpi]_{\rm dB}+[\gamma_{\rm b}]_{\rm dB})}{8.7 M}\right)\hfill\IEEEnonumber\\
&\overset{{\rm (iii)}}{\approx}&	\frac{N_{n_1}}{m(M+1)} {\rm cosh}^{-R}\left(\frac{n_1([\varpi]_{\rm dB}+[\gamma_{\rm b}]_{\rm dB})}{8.7 M}\right)\hfill
	\IEEEeqnarraynumspace
\end{IEEEeqnarray}
where $[x]_{\rm dB}=10\log_{10}(x)$ is the value of $x$ in decibel, and
\begin{IEEEeqnarray}{ccc}
	\label{Eq_XI}
\Xi(\gamma_{\rm b},M)=\sum_{k=1}^{K}\left(\frac{N_{n_k} {{\rm cosh}}{\left(\frac{n_k([\varpi]_{\rm dB}+[\gamma_{\rm b}]_{\rm dB})}{8.7 M}\right)}}{N_{n_1} {{\rm cosh}}{\left(\frac{n_1([\varpi]_{\rm dB}+[\gamma_{\rm b}]_{\rm dB})}{8.7 M}\right)}}\right)^{-R}.	
	\IEEEeqnarraynumspace
\end{IEEEeqnarray}
Note that in \eqref{Eq_delta_new3}, (i) is resulted from substituting $r\approx(\varpi \gamma_{\rm b})^\frac{1}{M}$  into \eqref{Eq_delta_new2}, and (ii) is obtained from the fact that $\ln(x)=\frac{\log_{10}(x)}{\log_{10}(e)}\approx\frac{\log_{10}(x)}{0.435}$. Furthermore, noting that $n_k>n_1$ for $k\geq2$, and ${\rm cosh}(x)$ is an increasing function of $x$, it is concluded from \eqref{Eq_XI} that $\Xi(\gamma_{\rm b},M)\approx 1$ when SNR is large. This fact leads to (iii) in \eqref{Eq_delta_new3}.}

Using the union bounding technique, it can be shown that the asymptotic BER of the system is proportional to $g_{\Phi}(d_{\min},\Psi,\xi)$, where $d_{\min}$ is the minimum free Hamming distance of the code \cite{Caire98}. More specifically,
	\begin{IEEEeqnarray}{ccc}
		\label{Eq_BER_union_bound}
		\log_{10}(P_{\rm UB}) &\;\approx\;&  \log_{10}(g_{\Phi}(d_{\min},\Psi,\xi))+\text{constant} \hfill
		\IEEEnonumber\\
			&\;\overset{(\rm i)}{\propto}\;& \Lambda(d_{\min},\Phi,M)-\Omega(R,d_{\min},\Phi,M,\gamma_{\rm b}),\hfill
		\IEEEeqnarraynumspace
	\end{IEEEeqnarray} 	
where (i) is resulted from substituting \eqref{Eq_delta_new3} into \eqref{Eq_average_PEP_tightset} in which
	\begin{IEEEeqnarray}{cc}
	\label{Eq_Lambda}
	\Lambda(d_{\min},\Phi,M)=d_{\min}\log_{10}\left(\frac{N_{n_1}}{m(M+1)}\right),
	\IEEEeqnarraynumspace
\end{IEEEeqnarray} 	
and
	\begin{IEEEeqnarray}{cc}
	\label{Eq_Omega}
	\Omega(R,d_{\min},\Phi,M,\gamma_{\rm b})=
	Rd_{\min}\log_{10}\left({\rm cosh}\left(\frac{n_1([\varpi]_{\rm dB}+[\gamma_{\rm b}]_{\rm dB})}{8.7 M}\right)\right).
	\IEEEeqnarraynumspace
\end{IEEEeqnarray}	

A few remarks regarding \eqref{Eq_BER_union_bound} are as follows.

\begin{remark} \label{remark2}
Recall that $n_1$ is the smallest value that $|l-\Phi(l,w)|$ takes for every $l\in \{0,\cdots,M\}$ and $w\in\{1,\cdots,m\}$. The function $\Phi(l,w)$ depends on (i) whether the FF or EFF scenario is considered: $\Phi(l,w)=\kappa(l,w)$ and $\Phi(l,w)=\varrho(l,w)$, respectively, and (ii) the applied mapping. Consequently, the functions $\Lambda(d_{\min},\Phi,M)$ and $\Omega(R,d_{\min},\Phi,M,\gamma_{\rm b})$ depend on $\Phi$ through $N_{n_1}$ and $n_1$, respectively.
\end{remark}

\begin{remark} \label{remark3}(System's diversity order). The diversity order of the system is given by
	\begin{IEEEeqnarray}{cc}
	\label{Eq__diversity}
	G_{\rm d}^{\rm sys}&=-\lim\limits_{\gamma_{\rm b}\rightarrow \infty} \frac{\log_{10}(P_{\rm UB})}{\log_{10}(\gamma_{\rm b})}\hfill=10 \lim\limits_{[\gamma_{\rm b}]_{\rm dB}\rightarrow \infty}\frac{\Omega(R,d_{\min},\Phi,M,\gamma_b)}{[\gamma_{\rm b}]_{\rm dB}}\hfill\IEEEnonumber\\
	&\overset{(\rm  i)}{=}10Rd_{\min}\lim\limits_{[\gamma_{\rm b}]_{\rm dB}\rightarrow \infty}\frac{\log_{10}\left({\rm cosh}\left(\frac{n_1[\gamma_{\rm b}]_{\rm dB}}{8.7 M}\right)\right)}{[\gamma_{\rm b}]_{\rm dB}}\hfill\IEEEnonumber\\
	&\overset{(\rm ii)}{=}\frac{10n_1Rd_{\min}}{8.7 M}\lim\limits_{u\rightarrow \infty}\frac{\log_{10}\left({\rm cosh}(u)\right)}{u}\overset{(\rm iii)}{=}\frac{n_1Rd_{\min}}{2 M},\hfill
	\IEEEeqnarraynumspace
\end{IEEEeqnarray}
where (i) is obtained by substituting in \eqref{Eq_Omega}, (ii) is obtained by the change of variable $u=\frac{n_1[\gamma_{\rm b}]_{\rm dB}}{8.7 M}$, and (iii) is due to the fact that $\lim\limits_{u\rightarrow \infty}\frac{\log_{10}\left({\rm cosh}(u)\right)}{u}=\frac{1}{\ln(10)}$. 

According to \eqref{Eq__diversity}, the system's diversity order scales with the number of antennas $R$, the minimum free Hamming distance $d_{\min}$ of the code, and the reciprocal of the modulation order $M$. Moreover, $G_{\rm d}^{\rm sys}$ is dependent on the case of FF or EFF, and the employed signal mapping through $n_1$ (see Remark \ref{remark2}). In general, a mapping that results in a higher value of $n_1$ is more desirable since it leads to a higher diversity order. However, increasing $n_1$ might not always be possible. In particular, for the case of FF, we have $\Phi(l,w)=\kappa(l,w)$. According to the definition of $\kappa(l,w)$, one can easily verify that for all $l\in\{0,\cdots,M\}$ and $w\in\{1,\cdots,m\}$, the smallest value of $j$ for which $|l-\kappa(l,w)|=j$ is always $j=1$. In other words, in the FF scenario (BICEM without iterative decoding) we always have $n_1=1$ regardless of the signal mapping. On the contrary, for the case of EFF, one can find mappings that result in $n_1>1$. It is important to note that, in contrast to our case, the diversity order in traditional BICM-ID schemes (e.g., \cite{Chindapol01,Tran06-Broad}) is independent of the signal mapping. The following example further clarifies this important point. 
\end{remark}
\begin{example} \label{Ex2}
Consider the Gray labeling of Example \ref{Ex1}. From Table \ref{table_EX1} one can verify that $n_1=1$ for both the FF and EFF cases. For example, we see that $|0-\kappa(0,3)|=1$ and $|1-\varrho(1,3)|=1$. Now, let us examine the mapping $\mathbf{L}_{1}=[0\ 5\ 6\ 3\ 4\ 1\ 2\ 7]^\top$. Table \ref{table_EX2} presents values of $\kappa(l,w)$ and $\varrho(l,w)$ for this mapping. It can be simply verified that $n_1=1$ for the FF scenario (e.g., $|3-\kappa(3,2)|=1$). Yet, the smallest value $j$ for which $|l-\varrho(l,w)|=j$ is $2$. Hence, $n_1=2$ for the case of EFF.  $\blacksquare$
\end{example}
\begin{table}[tb]
	\centering
	\caption{Functions $\kappa(l,w)$ and $\varrho(l,w)$ for signal mapping $\mathbf{L}_1$ considered in Example \ref{Ex2}.}
	\label{table_EX2}
	\begin{tabular}{|c|cc|c|c|c|c|cc|c}
		\cline{1-9}
		\backslashbox{$w$}{$l$} & \multicolumn{1}{c|}{0} & 1 & 2 & 3 & 4 & 5 & \multicolumn{1}{c|}{6} & 7 &                                              \\ \hline
		1  & \multicolumn{1}{c|}{1} & 0 & 3 & 2 & 5 & 4 & \multicolumn{1}{c|}{7} & 6 & \multicolumn{1}{c|}{\multirow{3}{*}{$\kappa(l,w)$}}  \\ \cline{1-9}
		2  & \multicolumn{2}{c|}{2}     & 1 & 4 & 3 & 6 & \multicolumn{2}{c|}{5}     & \multicolumn{1}{c|}{}                        \\ \cline{1-9}
		3  & \multicolumn{1}{c|}{1} & 0 & 3 & 2 & 5 & 4 & \multicolumn{1}{c|}{7} & 6 & \multicolumn{1}{c|}{}                        \\ \hline \hline
		1  & \multicolumn{1}{c|}{5} & 4 & 7 & 6 & 1 & 0 & \multicolumn{1}{c|}{3} & 2 & \multicolumn{1}{c|}{\multirow{3}{*}{$\varrho(l,w)$}} \\ \cline{1-9}
		2  & \multicolumn{1}{c|}{6} & 7 & 4 & 5 & 2 & 3 & \multicolumn{1}{c|}{0} & 1 & \multicolumn{1}{c|}{}                        \\ \cline{1-9}
		3  & \multicolumn{1}{c|}{4} & 5 & 6 & 7 & 0 & 1 & \multicolumn{1}{c|}{2} & 3 & \multicolumn{1}{c|}{}                        \\ \hline
	\end{tabular}
\end{table}

\begin{remark}\label{remark4}
	 When comparing two different mappings, one mapping might exhibit a coding gain over another mapping (in either BICEM or BICEM-ID system) by comparing the number $N_{n_1}$ of the nearest neighbors. To be more specific, suppose $N_{n_1}$ is equal to $\eta_1$ and $\alpha \eta_1$ for two different mappings $\textbf{L}_{i}$ and $\textbf{L}_{j}$, respectively, and $\alpha<1$. From \eqref{Eq_Lambda}, the value of $\Lambda(\cdot)$, which translates into a horizontal shift of the BER curve in the logarithm domain (see \eqref{Eq_BER_union_bound}) for the mapping $\textbf{L}_{j}$ is
		\begin{IEEEeqnarray}{cc}
		\label{Eq_Lambda_coding}
		\Lambda(d_{\min},\Phi,M)=d_{\min}\log_{10}\left(\frac{\alpha N_{n_1}}{m(M+1)}\right)
		=d_{\min}\log_{10}(\alpha)+d_{\min}\log_{10}\left(\frac{ N_{n_1}}{m(M+1)}\right).
		\IEEEeqnarraynumspace
	\end{IEEEeqnarray}
Hence, $\textbf{L}_{j}$ enjoys a coding gain of $-d_{\min}\log_{10}(\alpha)$ over $\textbf{L}_{i}$.
\end{remark}

The following example highlights the main points in Remarks \ref{remark2}--\ref{remark4}.
\begin{example}\label{Ex3}
	Consider the mapping rules in Examples \ref{Ex1} and \ref{Ex2}. From Tables \ref{table_EX1} and \ref{table_EX2}, one can easily obtain $N_1,\cdots,N_7$ for the Gray mapping (Example \ref{Ex1}) and the $\mathbf{L}_1$ mapping (Example \ref{Ex2}) for both the FF and EFF scenarios. These numbers are shown in Table \ref{table_EX3}. Recall that the performance in the FF mode (i.e., the BER of BICEM without iteration) also affects the performance of BICEM-ID. More specifically, a good signal mapping is the one that (i) results in a FF error rate that is small enough such that the first iteration works well, and (ii) leads to the smallest error rate in the EFF scenario. We can compare the two mappings in FF and EFF cases as follows.

For the FF case, the two signal mappings result in the same diversity order since $n_1=1$ for both. Yet, it is expected that the performance of the system that employs  $\mathbf{L}_1$ mapping is worse than the one with Gray mapping at all SNRs. This is due to the fact that $N_1=22$ for the former, which is larger than $N_1=16$ for the latter. In other words, the $8$-ary system of Example \ref{Ex1} enjoys a larger coding gain compared to the system of Example \ref{Ex2}.
		
For the EFF case, the $\mathbf{L}_1$ mapping significantly outperforms the Gray mapping. More specifically, since $n_1=2$ for the $\mathbf{L}_1$ mapping, its diversity order is twice that of the Gray mapping (with $n_1=1$). Moreover, $N_{n_1}=N_1=14$ for the Gray mapping, whereas $N_{n_1}=N_{2}=4$ for $\mathbf{L}_1$ mapping, which leads to a coding gain over the Gray labeling.
		
Finally, when considering the whole SNR range of interest, it is expected that in low-to-medium SNRs, the system that employs the Gray mapping works better than the one with the $\mathbf{L}_1$ mapping, thanks to its better FF performance. In contrast, for the high SNR region the system employing the $\mathbf{L}_1$ mapping can significantly outperform the system with the Gray mapping. $\blacksquare$
\end{example}

\begin{table}[tb]
	\centering
	\caption{Distribution of $n_k$ and $N_{n_k}$ for the Gray and $\mathbf{L}_1$ mappings in Examples \ref{Ex1} and \ref{Ex2} for both FF and EFF scenarios.}
	\label{table_EX3}
	\begin{tabular}{|c|c|c|c|c|c|c|c|c}
		\cline{1-8}
		\backslashbox{$\mathbf{L}$}{\\$N_k$} & $N_1$ & $N_2$ & $N_3$ & $N_4$ & $N_5$ & $N_6$ & $N_7$ &                                                                                            \\ \hline
		$ \mathbf{L}_1$           & $22$ & $2$  & $0$  & $0$  & $0$  & $0$  & $0$  & \multicolumn{1}{c|}{\multirow{2}{*}{\begin{tabular}[c]{@{}c@{}}FF\\ \end{tabular}}}  \\ \cline{1-8}
		$ {\rm Gray}$         & $16$ & $6$  & $2$  & $2$  & $0$  & $0$  & $0$  & \multicolumn{1}{c|}{}                                                                      \\ \hline  \hline
		$ \mathbf{L}_1$           & $0$  & $4$  & $4$  & $8$  & $4$  & $4$  & $0$  & \multicolumn{1}{c|}{\multirow{2}{*}{\begin{tabular}[c]{@{}c@{}}EFF\end{tabular}}} \\ \cline{1-8}
		$ {\rm Gray}$         & $14$ & $0$  & $6$  & $0$  & $2$  & $0$  & $2$  & \multicolumn{1}{c|}{}                                                                      \\ \hline
	\end{tabular}
\end{table}				

\begin{figure}[tb]
	\centering
	\includegraphics[width=4.25in]{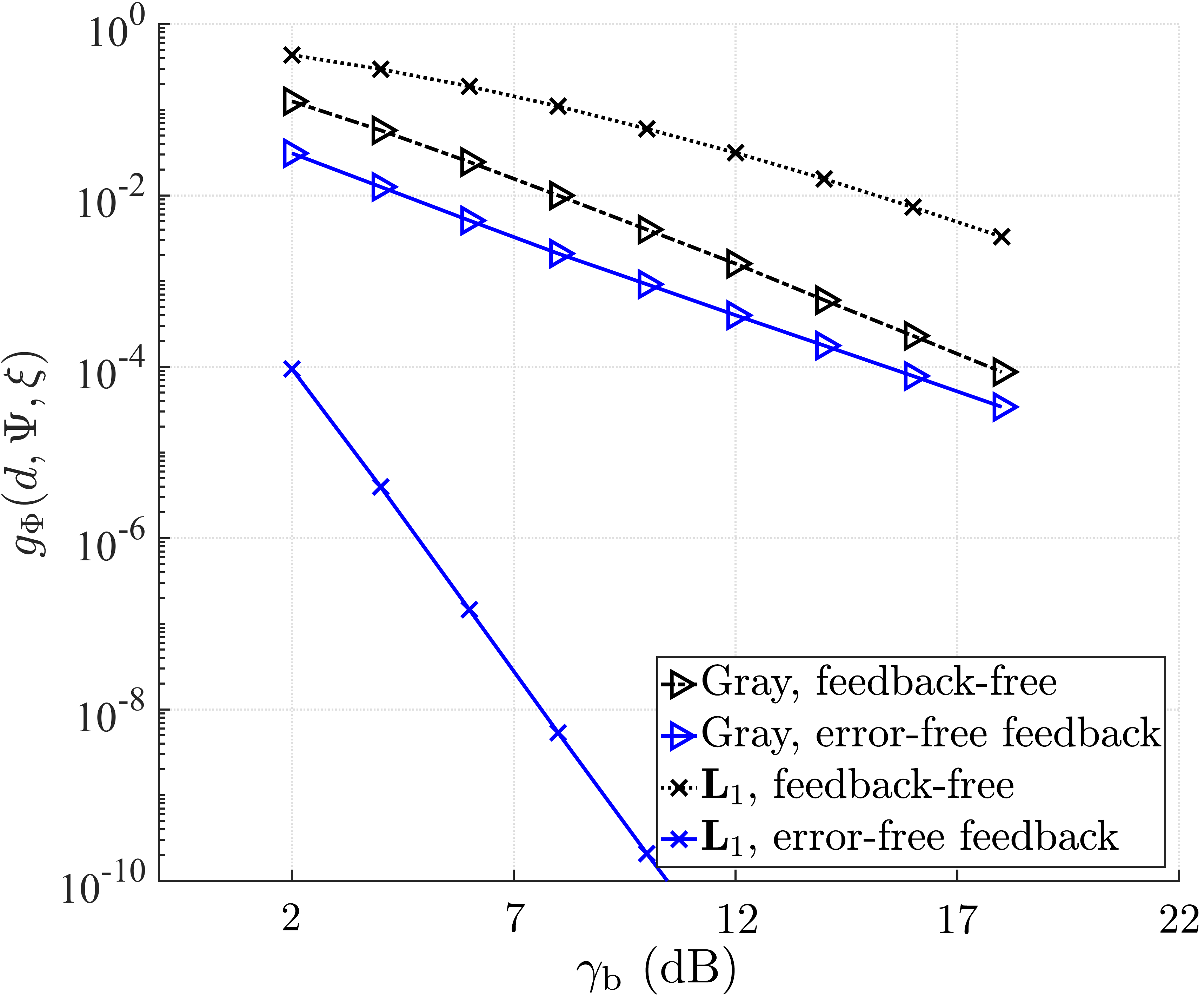}
	\caption{Upper-bounds on $g_\Phi(d,\Psi,\xi)$ for the $8$-ary systems in Examples \ref{Ex1} and \ref{Ex2} calculated from \eqref{Eq_average_PEP_tightset} and \eqref{Eq_delta}. Here the systems employ $R=5$ receive antennas and a convolutional code with $d_{\min}=10$.}\label{fig_Ex3}
\end{figure}

		 To have a clearer view of the performance of the $8$-ary systems studied in Examples \ref{Ex1} and \ref{Ex2}, upper-bounds on the average PEP $g_\Phi(d,\Psi,\xi)$ of both systems are plotted in Fig. \ref{fig_Ex3}. Here both systems employ $R=5$ receive antennas and a convolutional code having $d_{\min}=10$. The bounds are obtained by substituting \eqref{Eq_delta} into \eqref{Eq_average_PEP_tightset} and calculating values $|l-\Phi(l,w)|$ for both the FF and EFF scenarios. For the system with the $\mathbf{L}_1$ mapping in the EFF mode, from \eqref{Eq__diversity} and Table \ref{table_EX3} we expect that the error curve should have a slope of $\frac{Rn_1d_{\min}}{20M}=\frac{5\times 2\times 10}{20\times 7}=0.7143$, which can be verified from Fig. \ref{fig_Ex3} (the slope of the curve in Fig. \ref{fig_Ex3} is $\approx 0.712$). This shows that the approximations in \eqref{Eq_delta_new3}--\eqref{Eq__diversity} capture well the behavior of BICEM-ID formulated by \eqref{Eq_average_PEP_tightset} and \eqref{Eq_delta}. Moreover, from \eqref{Eq__diversity}, one expects that the three other curves (Gray mapping in the FF and EFF modes, and the $\mathbf{L}_1$ mapping in FF mode) have the same diversity order (because $n_1=1$ for all of them). This fact can be observed from the figure, in which the three curves have almost the same slope.
				
Finally, comparison of the curves in Fig. \ref{fig_Ex3} corroborates our remarks in Example \ref{Ex3}. Particularly, we note that in the FF scenario, the Gray mapping shows a better performance compared to $\mathbf{L}_1$ mapping, thanks to its coding gain. This means that the Gray mapping outperforms the  $\mathbf{L}_1$ mapping in the first stage of decoding.

\section{Signal Mappings for EB Constellations}\label{sec-alg}

		In the previous section we obtained bounds on the asymptotic BERs of BICEM and BICEM-ID. We then developed their approximations to reveal the impacts of different system parameters.  In this section we investigate the problem of signal mapping designs for the proposed system. For ease of exposition, we present the solution for an $8$-ary constellation ($m=3$, or $M=7$), although the same method can be generalized to an arbitrary $m$.
		
		We represent all $8!=40,320$ distinct mappings of an $8$-ary constellation as a $8\times 8!$ matrix $\underline{\mathbf{\Theta}}$ so that the $i$th column, $i=1,\cdots,8!$, of this matrix defines a specific labeling $\mathbf{L}$. Recall that, for fixed $\gamma_{\rm b}$, $R$, and $d_{\min}$, we can find two upper bounds for each mapping. These upper bounds are obtained from \eqref{Eq_average_PEP_tightset} by computing $\delta_{\Phi}(\Psi,\xi)$ in \eqref{Eq_delta} for $\Phi(l,w)=\kappa(l,w)$ and $\Phi(l,w)=\varrho(l,w)$. This results in two vectors, $\mathbf{\Delta}_{\kappa}$ and $\mathbf{\Delta}_{\varrho}$, each of length $8!$. Particularly, $[\mathbf{\Delta}_{\kappa}]_i$ and $[\mathbf{\Delta}_{\varrho}]_i$ are $g_{\kappa}(d,\Psi,\xi)$ and $g_{\varrho}(d,\Psi,\xi)$ for the FF and EFF scenarios, respectively, when the signal mapping in the $i$th column of $\underline{\mathbf{\Theta}}$ is employed.

Consider a permutation matrix $\underline{\mathbf{\Pi}}_{\kappa}$, which sorts elements of $\mathbf{\Delta}_{\kappa}$ in an ascending order and produces $\mathbf{\Delta}_{\kappa,{\rm sort}}=\underline{\mathbf{\Pi}}_{\kappa}\mathbf{\Delta}_{\kappa}$. This same matrix permutes elements of $\mathbf{\Delta}_{\varrho}$ and columns of $\underline{\mathbf{\Theta}}$ to produce $\mathbf{\Delta}_{\varrho,{\rm sort}}=\underline{\mathbf{\Pi}}_{\kappa}\mathbf{\Delta}_{\varrho}$ and $\underline{\mathbf{\Theta}}_{\rm sort}=\underline{\mathbf{\Theta}}\underline{\mathbf{\Pi}}_{\kappa}^{\top}$. The mapping represented by the $i$th column of $\underline{\mathbf{\Theta}}_{\rm sort}$ shall be named $\mathbf{L}_i$. Next, we partition the vector $\mathbf{\Delta}_{\kappa,{\rm sort}}$ into $p$ sub-vectors $\mathbf{d}_{\rho,\kappa}$, $\rho=1,\cdots,p$, i.e., $\mathbf{\Delta}_{\kappa,{\rm sort}}=[\mathbf{d}_{1,\kappa}^{\top},\mathbf{d}_{2,\kappa}^{\top},\cdots,\mathbf{d}_{p,\kappa}^{\top}]^{\top}$. The length $\mu_{\rho}$ of $\mathbf{d}_{\rho,\kappa}$ is constrained by parameter $\epsilon$, which is the maximum absolute logarithmic difference across elements of $\mathbf{d}_{\rho,\kappa}$. More specifically, since $\mathbf{\Delta}_{\kappa,{\rm sort}}$, and hence sub-vectors $\mathbf{d}_{\rho,\kappa}$ include elements in an ascending order, $\mu_{\rho}$ is the maximum value satisfying
		\begin{IEEEeqnarray}{ll}
			\label{Eq_partiton_condition}		 \log_{10}\left([\mathbf{d}_{\rho,\kappa}]_{\mu_{\rho}}\right)-\log_{10}\left([\mathbf{d}_{\rho,\kappa}]_1\right)\leq \epsilon, \ \  \rho=1,\cdots,p.
			\IEEEeqnarraynumspace
		\end{IEEEeqnarray}

The constraint of \eqref{Eq_partiton_condition} ensures that the error values included in each sub-vector are $\epsilon$-close to each other on the logarithmic scale. Hence, by choosing $\epsilon$ small enough, one can say that the elements in every sub-vector $\mathbf{d}_{\rho,\kappa}$ are approximately equal. We then partition
		$\mathbf{\Delta}_{\varrho,{\rm sort}}$ and $\underline{\mathbf{\Theta}}_{\rm sort}$ into $p$ groups based on the partitioning of $\mathbf{\Delta}_{\kappa,{\rm sort}}$. In other words, $\mathbf{\Delta}_{\varrho,{\rm sort}}=[\mathbf{d}_{1,\varrho}^{\top},\mathbf{d}_{2,\varrho}^{\top},\cdots,\mathbf{d}_{p,\varrho}^{\top}]^{\top}$, such that $\mathbf{d}_{\rho,\varrho}$ is a length-$\mu_\rho$ vector. Similarly, $\underline{\mathbf{\Theta}}_{\rm sort}=[\underline{\mathbf{L}}_1,\cdots,\underline{\mathbf{L}}_p]$, where $\underline{\mathbf{L}}_\rho$ is constructed from $\mu_\rho$ column vectors. Now, we define the best mapping within partition $\rho$, $1\leq \rho\leq p$ as follows.
		\begin{definition}
		 \label{defin1}
			Consider the $\rho$th partition of $\underline{\mathbf{\Theta}}_{\rm sort}$, denoted by $\underline{\mathbf{L}}_\rho=[\mathbf{L}_{g_\rho+1},\cdots,\mathbf{L}_{g_\rho+\mu_\rho}]$, where $1\leq \rho\leq p$, $g_1=0$, and $g_\rho=\sum_{t=1}^{\rho-1}\mu_t$ for $\rho\neq1$.
			The mapping vector $\mathbf{L}_{g_\rho+l^*_\rho}$ is the best mapping in partition $\rho$ if
				\begin{IEEEeqnarray}{ll}
				\label{Eq_best_mapping}
				l_\rho^*=\argmin_{l\in \{1,\cdots,\mu_\rho\}} [\mathbf{d}_{\rho,\varrho}]_{l}.
				\IEEEeqnarraynumspace
			\end{IEEEeqnarray}
		$\blacksquare$
		\end{definition}
		
Let us define $\underline{\mathbf{\Theta}}_{\rm best}=[\mathbf{L}_{g_1+l_1^*},\cdots,\mathbf{L}_{g_p+l_p^*}]$ as the set of best mappings in all $p$ partitions. Corresponding to $\underline{\mathbf{\Theta}}_{\rm best}$, we form $\mathbf{\Delta}_{\kappa,{\rm best}}$ and $\mathbf{\Delta}_{\varrho,{\rm best}}$. In this regard, $[\mathbf{\Delta}_{\kappa,{\rm best}}]_\rho=[\mathbf{d}_{\rho,\kappa}]_{l_\rho^*}$ and $[\mathbf{\Delta}_{\varrho,{\rm best}}]_\rho=[\mathbf{d}_{\rho,\varrho}]_{l_\rho^*}$ are the error values corresponding to the best mappings in the $\rho$th partition in the FF and EFF scenarios. The following remarks explain the motivation behind Definition \ref{defin1} and provide more details on the construction of $\underline{\mathbf{\Theta}}_{\rm best}$.

\begin{remark}
	As mentioned before, the condition in \eqref{Eq_partiton_condition} ensures that the error values in each partition of $\mathbf{\Delta}_{\kappa,{\rm sort}}$ (vector of errors in the FF mode) are almost equal. Obviously, among all mappings that result in the same error in the FF case, the best mapping is the one that minimizes the error rate in the EFF scenario. This mapping is found by \eqref{Eq_best_mapping} in Definition \ref{defin1}.
\end{remark}
\begin{remark}
	When $\epsilon\rightarrow 0$, the error values in $\mathbf{d}_{\rho,\kappa}$, $1\leq \rho\leq p$, get closer to each other (the difference between the maximum and minimum values becomes smaller). In practice, for some $0<\epsilon<\epsilon_0$ all values in $\mathbf{d}_{\rho,\kappa}$ will be equal. Clearly, by decreasing $\epsilon$ one expects that the number $p$ of partitions increases. Yet, it is worth noting that for any $\epsilon$, $p$ is always significantly smaller than the number $8!$ of all permutations. This roots in the fact that for each specific mapping, there exists a large number of mappings resulting in the same value of FF error.
\end{remark}

Algorithm \ref{Alg1} presents a pseudo-code for constructing the matrix of best mappings for a general modulation order $2^m$.

\begin{example} \label{EX4}
Consider an $8$-ary BICEM-ID system employing $R=5$ receive antennas, and a rate--$\frac{2}{3}$ code of $d_{\min}=10$. Table \ref{table_EX4} lists the set of best mappings. These mappings, which are columns of $\underline{\mathbf{\Theta}}_{\rm best}$, i.e., $\mathbf{L}_{g_\rho+l_\rho^*}$, $\rho=1,\cdots,14$, are obtained from Algorithm \ref{Alg1} for $\epsilon=4\times 10^{-4}$ and at $\gamma_{\rm b}$ between $9.5$ to $14.5$ (dB). The values $[\mathbf{\Delta}_{\kappa,{\rm best}}]_\rho$ and $[\mathbf{\Delta}_{\varrho,{\rm best}}]_\rho$ (i.e., the upper bounds on the average PEP in the FF and EFF modes, respectively for the $\rho$th mapping) are also given on base-$10$ logarithmic scale. It is observed from the table that at every SNR, $[\mathbf{\Delta}_{\kappa,{\rm best}}]_\rho$ increases with $\rho$, whereas $[\mathbf{\Delta}_{\varrho,{\rm best}}]_\rho$ shows an opposite behavior. Consequently, the labels with a lower index $\rho$ show a better performance in the FF scenario (i.e., BICEM without any iterations in the receiver). On the contrary, as we move toward the bottom of the table, the mappings yield better error rates in the EFF case. Depending on the working SNR, one of the mappings in Table \ref{table_EX4} can be preferable for BICEM-ID. These observations are elaborated in more details in the next section. $\blacksquare$
\end{example}

\begin{algorithm}[tb]
	\caption{Best mappings for $2^m$-ary BICEM-ID.} \label{Alg1}
	\begin{algorithmic}[1]
		\renewcommand{\algorithmicrequire}{\textbf{Input:}}
		\renewcommand{\algorithmicensure}{\textbf{Output:}}
		\REQUIRE $\underline{\mathbf{\Theta}}$, $\gamma_{\rm b}, R, d_{\min}, m$, $\epsilon$
		\ENSURE  $\underline{\mathbf{\Theta}}_{\rm best}$
		\\	\textit{Initialize} :  $i=0$, $\rho=0$, $\mu_0=0$, $g_0=0$, $z=1$
		\WHILE {$i< (2^m)!$}
		\STATE $i\leftarrow i+1$
		\STATE Use \eqref{Eq_delta} to obtain\\ $[\mathbf{\Delta}_\kappa]_i=\delta^{d_{\min}}(\Psi,\xi:[\underline{\mathbf{\Theta}}]_{:,i},\Phi=\kappa)$\\
		 $[\mathbf{\Delta}_\varrho]_i=\delta^{d_{\min}}(\Psi,\xi:[\underline{\mathbf{\Theta}}]_{:,i},\Phi=\varrho)$
		\ENDWHILE
		\STATE Find matrix $\underline{\mathbf{\Pi}}_{\kappa}$ that sorts  $\mathbf{\Delta}_\kappa$ in ascending order
		\STATE $\mathbf{\Delta}_{\kappa,{\rm sort}}=\underline{\mathbf{\Pi}}_{\kappa}\mathbf{\Delta}_{\kappa}$,\ \ \ $\mathbf{\Delta}_{\varrho,{\rm sort}}=\underline{\mathbf{\Pi}}_{\kappa}\mathbf{\Delta}_{\varrho}$,
		\ \ $\underline{\mathbf{\Theta}}_{\rm sort}=\underline{\mathbf{\Theta}}\underline{\mathbf{\Pi}}_{\kappa}^{\top}$
		\WHILE {$\mu_\rho< (2^m)!$}
		\STATE $\rho\leftarrow \rho+1$
		\STATE $g_\rho= g_{\rho-1}+\mu_{\rho-1}$
		\STATE Maximize $\mu_\rho$ such that $\frac{[\mathbf{\Delta}_{\kappa,{\rm sort}}]_{g_\rho+\mu_\rho}}{[\mathbf{\Delta}_{\kappa,{\rm sort}}]_{g_\rho+1}}\leq 10^\epsilon$
		\STATE Find $l_\rho^*=\argmin_{l\in \{1,\cdots, \mu_\rho\}} [\mathbf{\Delta}_{\varrho,{\rm sort}}]_{g_\rho+l}$
		\IF{$\rho=1$}
		\STATE $[\underline{\mathbf{\Theta}}_{\rm best}]_{:,1}=[\underline{\mathbf{\Theta}}_{\rm sort}]_{g_1+l_1^*}$
		\ELSIF{$[\mathbf{\Delta}_{\varrho,{\rm sort}}]_{g_\rho+l_\rho^*}<[\mathbf{\Delta}_{\varrho,{\rm sort}}]_{g_{\rho-1}+l_{\rho-1}^*}$}
		\STATE $z\leftarrow z+1$
		\STATE $[\underline{\mathbf{\Theta}}_{\rm best}]_{:,z}=[\underline{\mathbf{\Theta}}_{\rm sort}]_{g_\rho+l_\rho^*}$
		\ENDIF
		\ENDWHILE
		\STATE Return $\underline{\mathbf{\Theta}}_{\rm best}$
	\end{algorithmic}
\end{algorithm}

\begin{table*}[tb]
	\centering
	\caption{Table of best mappings for a $8$-ary BICEM-ID employing $R=5$ receive antennas, and a rate--$\frac{2}{3}$ convolutional code of minimum free Hamming distance $d_{\min}=10$. Note that $[\mathbf{\Delta}_{\kappa,{\rm best}}]_\rho$ and $[\mathbf{\Delta}_{\varrho,{\rm best}}]_\rho$ are written in decibel.}
	\label{table_EX4}
		\resizebox{\textwidth}{!}{
	\begin{tabular}{|c||c||cc||cc||cc||cc|}
		\hline
		\multirow{2}{*}{$\rho$} & \multirow{2}{*}{$[\underline{\mathbf{\Theta}}_{\rm best}]_{:,\rho}=\mathbf{L}_{g_\rho+l_\rho^*}$}                            & \multicolumn{2}{c||}{$\gamma_{\rm b}=9.5$ dB }                         & \multicolumn{2}{c||}{$\gamma_{\rm b}=11.5$ dB}                         & \multicolumn{2}{c||}{$\gamma_{\rm b}=13.5$ dB}                         & \multicolumn{2}{c|}{$\gamma_{\rm b}=14.5$ dB}                         \\ \cline{3-10}
		&                                                   & \multicolumn{1}{c|}{$[\mathbf{\Delta}_{\kappa,{\rm best}}]_\rho$} & $[\mathbf{\Delta}_{\varrho,{\rm best}}]_\rho$ & \multicolumn{1}{c|}{$[\mathbf{\Delta}_{\kappa,{\rm best}}]_\rho$} & $[\mathbf{\Delta}_{\varrho,{\rm best}}]_\rho$ & \multicolumn{1}{c|}{$[\mathbf{\Delta}_{\kappa,{\rm best}}]_\rho$} & $[\mathbf{\Delta}_{\varrho,{\rm best}}]_\rho$ & \multicolumn{1}{c|}{$[\mathbf{\Delta}_{\kappa,{\rm best}}]_\rho$} & $[\mathbf{\Delta}_{\varrho,{\rm best}}]_\rho$ \\ \hline
		$1$                  & {$[1  \   3  \   0 \    2 \    4 \    6 \    5\     7]^{\top}$} & \multicolumn{1}{c|}{$-2.3$}         & $-3.2$          & \multicolumn{1}{c|}{$-2.7$}         & $-3.74$         & \multicolumn{1}{c|}{$-3.1$}         & $-4.3$          & \multicolumn{1}{c|}{$-3.3$}         & $-4.57$         \\ \hline
		$2$                 & {$[1  \   3  \   0 \    2  \   4  \   5 \    6\     7]^{\top}$} & \multicolumn{1}{c|}{$-2.2$}         & $-3.4$          & \multicolumn{1}{c|}{$-2.6$}         & $-3.94$         & \multicolumn{1}{c|}{$-2.95$}        & $-4.5$          & \multicolumn{1}{c|}{$-3.15$}        & $-4.76$         \\ \hline
		$3$                  & {$[0    \ 3\     1 \    2\     4 \    6 \    5 \    7]^{\top}$} & \multicolumn{1}{c|}{$-2.1$}         & $-4$            & \multicolumn{1}{c|}{$-2.4$}         & $-4.6$          & \multicolumn{1}{c|}{$-2.8$}         & $-5.25$         & \multicolumn{1}{c|}{$-3$}           & $-5.55$         \\ \hline
		$4$                  & {$[0    \ 3     \ 1    \ 2   \  4   \  5    \ 6   \  7]^{\top}$} & \multicolumn{1}{c|}{$-2$}           & $-4.14$         & \multicolumn{1}{c|}{$-2.3$}         & $-4.8$          & \multicolumn{1}{c|}{$-2.65$}        & $-5.4$          & \multicolumn{1}{c|}{$-2.83$}        & $-5.74$         \\ \hline
		$5$                  & {$[5 \    1  \   4\     2  \   0  \   3 \    6 \    7]^{\top}$} & \multicolumn{1}{c|}{$-1.86$}        & $-4.25$         & \multicolumn{1}{c|}{$-2.2$}         & $-4.9$          & \multicolumn{1}{c|}{$-2.63$}        & $-5.5$          & \multicolumn{1}{c|}{$-2.82$}        & $-5.8$          \\ \hline
		$6$                  & {$[1\     2 \    0 \    3\     5\     6\     4 \    7]^{\top}$} & \multicolumn{1}{c|}{$-1.85$}        & $-5$            & \multicolumn{1}{c|}{$-2.18$}        & $-5.7$          & \multicolumn{1}{c|}{$-2.5$}         & $-6.48$         & \multicolumn{1}{c|}{$-2.68$}        & $-6.8$          \\ \hline
		$7$                  & ${[1 \    4    \ 5   \  6 \    0 \    3   \  2\     7]^{\top}}$ & \multicolumn{1}{c|}{$-1.65$}        & $-5.1$          & \multicolumn{1}{c|}{$-2$}           & $-5.8$          & \multicolumn{1}{c|}{$-2.35$}        & $-6.52$         & \multicolumn{1}{c|}{$-2.5$}         & $-6.86$         \\ \hline
		$8$                  & {$[4\     1  \   2   \  3 \    0  \   5  \   6\     7]^{\top}$} & \multicolumn{1}{c|}{$-1.6$}         & $-5.7$          & \multicolumn{1}{c|}{$-1.9$}         & $-6.4$          & \multicolumn{1}{c|}{$-2.3$}         & $-7$            & \multicolumn{1}{c|}{$-2.47$}        & $-7.4$          \\ \hline
		$9$                  & {$[1\     4  \   2  \   3  \   0 \    5  \   6  \   7]^{\top}$} & \multicolumn{1}{c|}{$-1.5$}         & $-6$            & \multicolumn{1}{c|}{$-1.8$}         & $-6.77$         & \multicolumn{1}{c|}{$-2.15$}        & $-7.45$         & \multicolumn{1}{c|}{$-2.33$}        & $-7.77$         \\ \hline
		$10$                & {$[4 \    2 \    1  \   3\     5  \   0 \    6\     7]^{\top}$} & \multicolumn{1}{c|}{$-1.42$}        & $-6.1$          & \multicolumn{1}{c|}{$-1.75$}        & $-6.82$         & \multicolumn{1}{c|}{$-2.1$}         & $-7.47$         & \multicolumn{1}{c|}{$-2.27$}        & $-7.8$          \\ \hline
		$11 $                & {$[1 \    2  \   4 \    3  \   0 \    5  \   6\     7]^{\top}$} & \multicolumn{1}{c|}{$-1.39$}        & $-6.4$          & \multicolumn{1}{c|}{$-1.7$}         & $-7.4$          & \multicolumn{1}{c|}{$-2$}           & $-8.3$          & \multicolumn{1}{c|}{$-2.2$}         & $-8.7$          \\ \hline
		$12 $                & {$[0\     6  \   5  \   3  \   1  \   2  \   4 \    7]^{\top}$} & \multicolumn{1}{c|}{$-1.33$}        & $-8$            & \multicolumn{1}{c|}{$-1.64$}        & $-8.85$         & \multicolumn{1}{c|}{$-1.97$}        & $-9.6$          & \multicolumn{1}{c|}{$-2.14$}        & $-9.95$         \\ \hline
		$13  $               & {$[0\     6 \    3   \  5   \  1   \  2  \   4 \    7]^{\top}$} & \multicolumn{1}{c|}{$-1.24$}        & $-8.4$          & \multicolumn{1}{c|}{$-1.5$}         & $-9.3$          & \multicolumn{1}{c|}{$-1.85$}        & $-10.1$         & \multicolumn{1}{c|}{$-2$}           & $-10.5$         \\ \hline
		$14 $                & {$[0\ 5\ 6 \ 3 \ 4\ 1\ 2\ 7]^{\top}$}                             & \multicolumn{1}{c|}{$-1.15$}        & $-9.3$          & \multicolumn{1}{c|}{$-1.43$}        & $-10.7$         & \multicolumn{1}{c|}{$-1.7$}         & $-12$          & \multicolumn{1}{c|}{$-1.9$}         & $-12.7$         \\ \hline
	\end{tabular}}
\end{table*}

\section{Simulation Results}\label{sec-num}

The key results regarding performance of BICEM-ID were established in the previous sections by theoretical analysis. In this section, we present simulation results to highlight and corroborate the analytical findings.

When an $8$-ary constellation is employed, the optimal $(3, 2, 10)$ convolutional code having the minimum free Hamming distance $d_{\min}=10$ (see \cite{Book-Shulin}) is selected. On the other hand, the optimal $(2,1,6)$ convolutional code having $d_{\min}=10$ is used when a $4$-ary constellation is considered. The information block length is $L_{\rm I}=6,000-\nu_{\rm c}$ bits, which means that the lengths of the random interleavers are $9,000$ and $12,000$ for $8$-ary and $4$-ary systems, respectively. Unless clearly mentioned, the number of iterations for iterative receivers is set to $8$. Results are compared with uncoded systems of the same information rate. As such, $8$-ary BICEM-ID systems (which use rate--$2/3$ code) are compared with the uncoded $4$-ary non-coherent EBM system. Similarly, the rate--$1/2$, $4$-ary coded modulation system is compared with the uncoded OOK system. Note that, in obtaining the BERs for the uncoded systems, Gray mapping is employed. 
Finally, the effect of the code constraint length $\nu_{\rm c}$ on the spectral efficiency is neglected as it is insignificant compared to the block length of the input bit vector.

\begin{figure}[tb]
	\centering
	\includegraphics[width=4.25in]{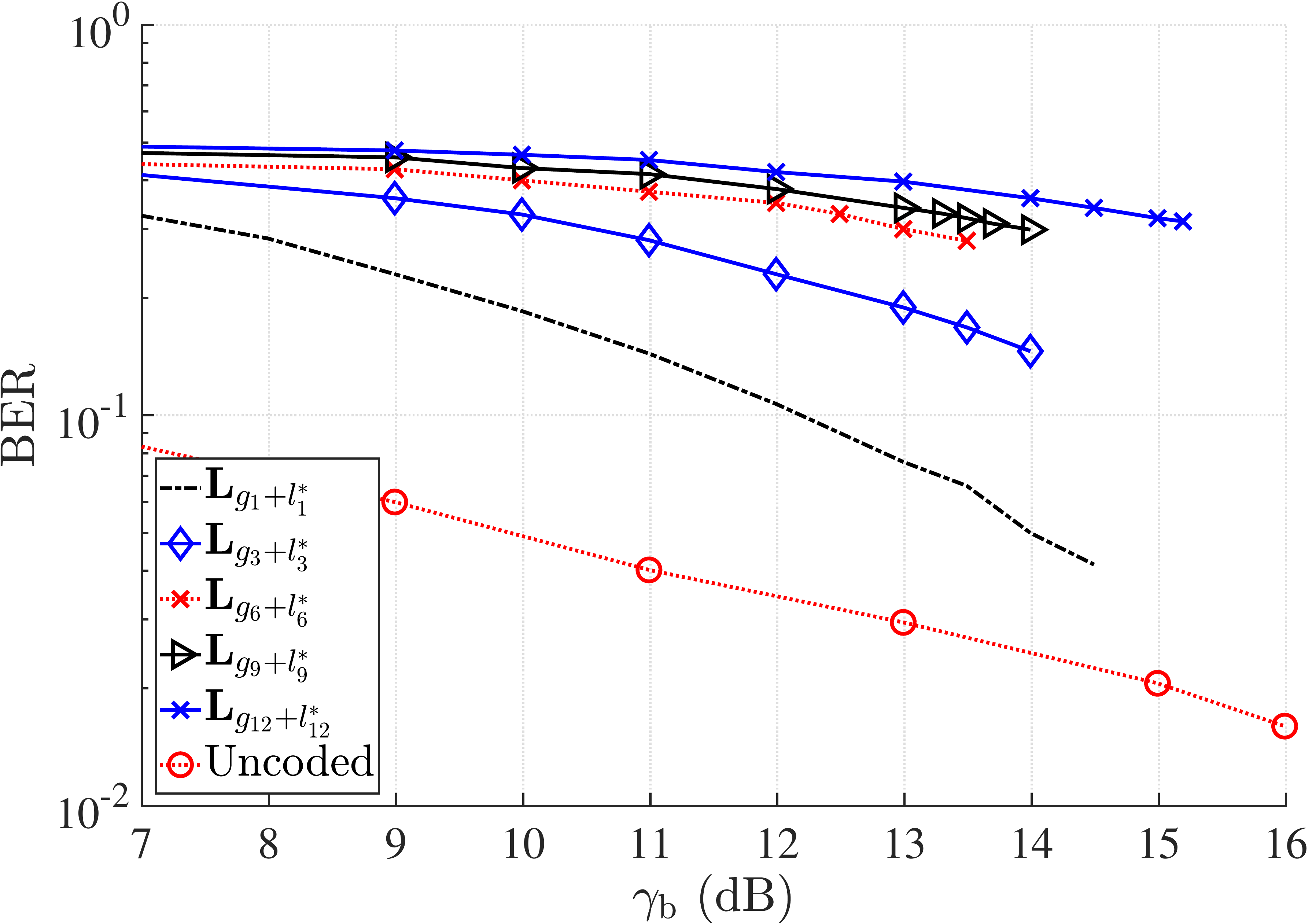}
	\caption{BERs of $8$-ary BICEM with five different mappings in Table \ref{table_EX4} and uncoded EBM (having the same spectral efficiency of $2$ bits/channel use).}
	\label{fig_map_comp}
\end{figure}

Figure \ref{fig_map_comp} compares the BERs of $8$-ary BICEM systems when five different mappings are employed. The five mappings are taken from Table \ref{table_EX4}, which were obtained with Algorithm \ref{Alg1}. These five mappings are specified as $\mathbf{L}_{g_1+l_1^*}$, $\mathbf{L}_{g_3+l_3^*}$, $\mathbf{L}_{g_6+l_6^*}$, $\mathbf{L}_{g_9+l_9^*}$, and $\mathbf{L}_{g_{12}+l_{12}^*}$ in the Table. For example, $\mathbf{L}_{g_{3}+l_{3}^*}$ is $[0\  3\ 1\ 2\ 4\ 6\ 5\ 7]^{\top}$. It is also assumed that $R=5$ antennas are used at the receiver. It can be observed that, for $\rho_1<\rho_2$, the mapping $\mathbf{L}_{g_{\rho_1}+l_{\rho_1}^*}$ outperforms $\mathbf{L}_{g_{\rho_2}+l_{\rho_2}^*}$. This fact was already shown analytically before with $[\mathbf{\Delta}_{\kappa,{\rm best}}]_{\rho_1}<[\mathbf{\Delta}_{\kappa,{\rm best}}]_{\rho_2}$ (see Table \ref{table_EX4} and algorithm \ref{Alg1}). Also plotted in Fig. \ref{fig_map_comp} is the BER curve for the $4$-ary uncoded non-coherent EBM system. As can be seen, at low-to-medium SNRs (SNRs below $16$ dB), there is no performance advantage provided by BICEM as compared to the uncoded system of the same spectral efficiency.

\begin{figure}[tb]
	\centering
	\includegraphics[width=4.25in]{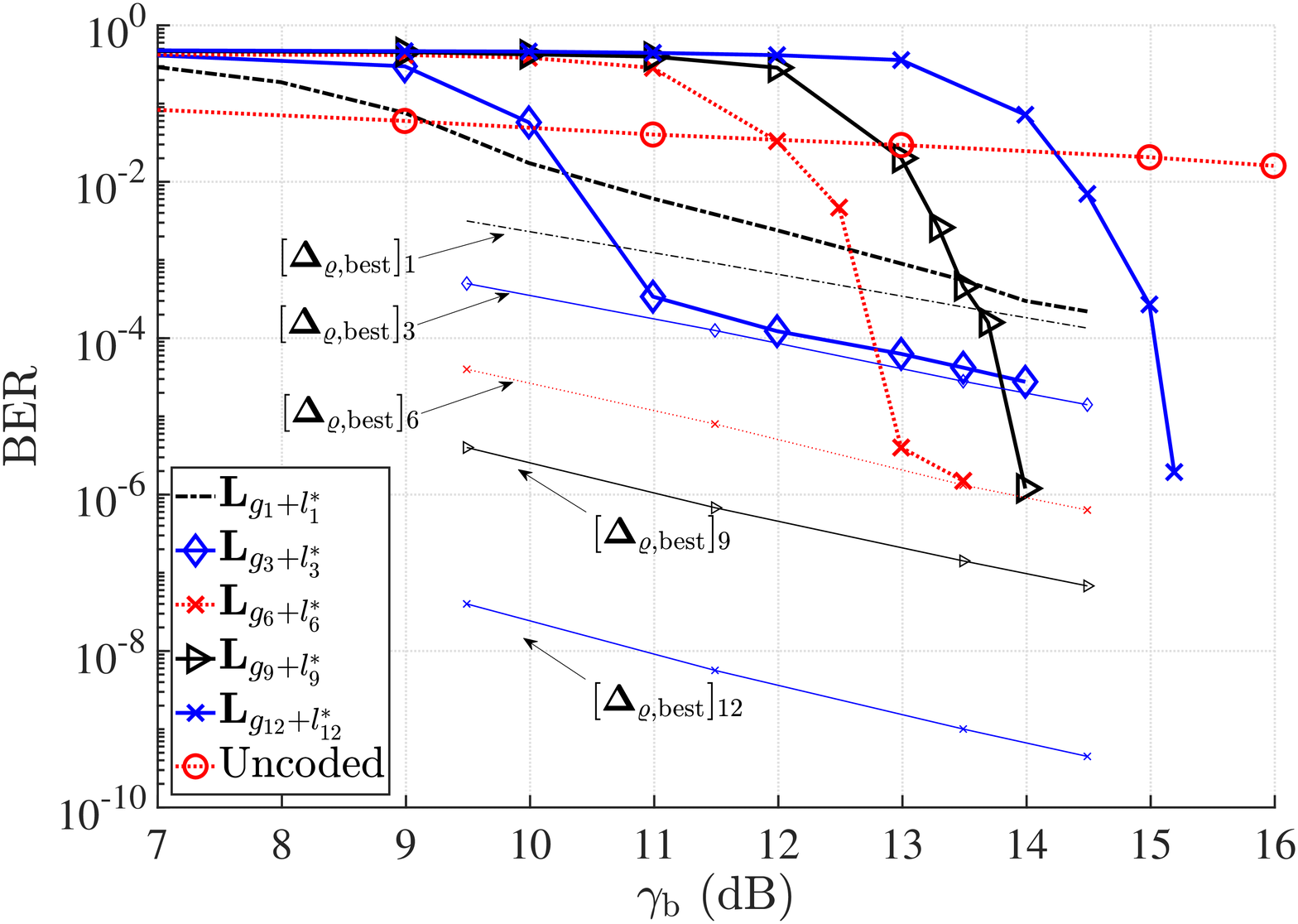}
	\caption{BERs of $8$-ary BICEM-ID with five different mappings in Table \ref{table_EX4} and uncoded EBM (having the same spectral efficiency of $2$ bits/channel use).}
	\label{fig_map_comp2}
\end{figure}

The behavior is very different in Fig. \ref{fig_map_comp2} in which the BER curves are shown for BICEM-ID systems of the same settings as those in Fig. \ref{fig_map_comp}, but obtained with $8$ iterations of the iterative receiver. Here, we can observe that the proposed BICEM-ID systems significantly outperform the uncoded system when the SNR is large enough. Specifically, the performance gain starts at $\gamma_{\rm b}\geq 9$ dB, $\gamma_{\rm b}\geq 10$ dB, $\gamma_{\rm b}\geq 12$ dB, $\gamma_{\rm b}\geq 13$ dB, and $\gamma_{\rm b}\geq 14$ dB, when $\mathbf{L}_{g_1+l_1^*}$, $\mathbf{L}_{g_3+l_3^*}$, $\mathbf{L}_{g_6+l_6^*}$, $\mathbf{L}_{g_9+l_9^*}$, and $\mathbf{L}_{g_{12}+l_{12}^*}$ mappings are employed, respectively. Another point worth mentioning in Fig. \ref{fig_map_comp2} is that, for $\rho_1<\rho_2$, the BICEM-ID system that applies the mapping $\mathbf{L}_{g_{\rho_1}+l_{\rho_1}^*}$ reaches its asymptotic performance at a lower SNR when compared to the system that applies the $\mathbf{L}_{g_{\rho_2}+l_{\rho_2}^*}$  mapping. Yet, the asymptotic performance of $\mathbf{L}_{g_{\rho_2}+l_{\rho_2}^*}$ is remarkably better than that of $\mathbf{L}_{q_{\rho_1}+l_{\rho_1}^*}$. This behavior is also expected, according to the results in Section \ref{sec-alg}. More specifically, as can be observed from Table \ref{table_EX4}, by increasing $\rho$, the error performance of $\mathbf{L}_{g_{\rho}+l_{\rho}^*}$ in the EFF mode improves, while that in the FF case degrades. As previously explained (see Example \ref{Ex3}), between two mappings, the one whose performance in the FF mode is better, reaches its error floor at a lower SNR. On the other hand, the better mapping, in terms of EFF performance, has a smaller error floor. As such, by  increasing $\rho$, we expect that $\mathbf{L}_{g_{\rho}+l_{\rho}^*}$ demonstrates a worse BER at low-to-midum SNRs, but a better asymptotic BER.

In Fig. \ref{fig_map_comp2} we have also plotted the asymptotic lines (indicated by $[\mathbf{\Delta}_{\varrho,{\rm best}}]_\rho$ with arrows in the figure), which are the upper bound obtained in Table \ref{table_EX4} for the EFF scenario. It is seen that for the mappings $\mathbf{L}_{g_1+l_1^*}$, $\mathbf{L}_{g_3+l_3^*}$, and $\mathbf{L}_{g_6+l_6^*}$, their BER curves indeed approach their asymptotic performance lines. On the other hand, as the BERs in the EFF mode are too low for the $\mathbf{L}_{g_9+g_9^*}$ and $\mathbf{L}_{g_{12}+l_{12}^*}$ mappings, such asymptotic performance behavior cannot be seen with the limited simulation results.

Comparing the BER curves of BICEM-ID under different mappings explains why a list of mappings (produced by Algorithm \ref{Alg1}) is a set of best mappings. Indeed, based on the working SNR and the objective BER, each of the mappings could be the best choice. For example, when the objective error rate is of order $10^{-6}$, employing the mapping $\mathbf{L}_{g_6+l_6^*}$ appears to be the best choice. This is because all other mappings reach this error rate at higher SNRs than that of the $\mathbf{L}_{g_6+l_6^*}$ mapping.

\begin{figure}[tb]
	\centering
	\includegraphics[width=4.25in]{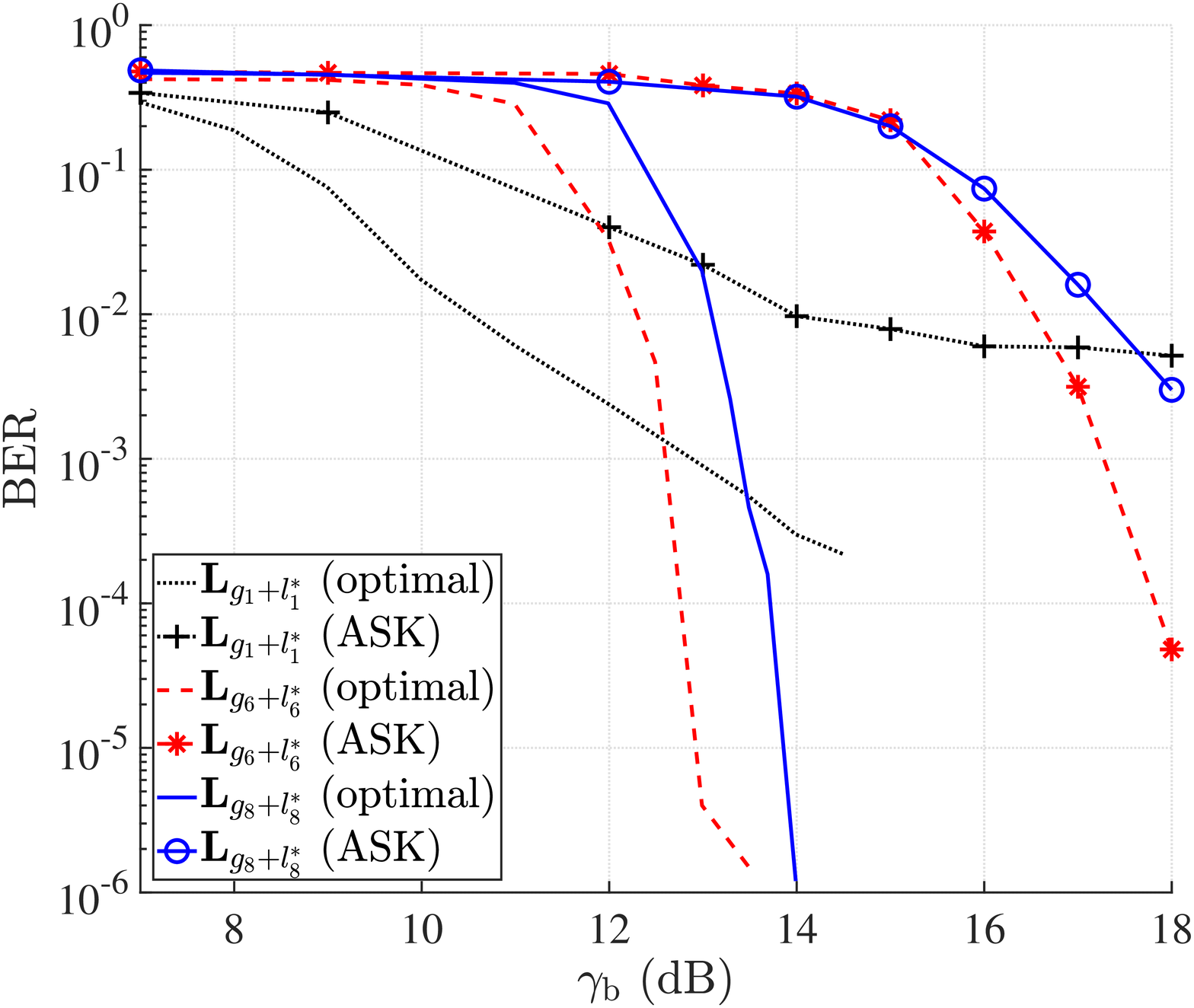}
	\caption{BER comparison of BICEM-ID under $8$-ary optimal EBM and ASK modulation.}
	\label{fig_map_rev23}
\end{figure}

The geometric distancing of signal points, resulted from optimal constellation design (see \eqref{Eq_all_one_poly} and Example \ref{Ex1} for more details) plays an important role in the efficiency of BICEM-ID. System performance under conventional ASK, which is an EBM with equidistant constellation points is evaluated in Fig. \ref{fig_map_rev23}. Significant performance gap between $8$-ary ASK (sub-optimal) and optimal EBM can be observed for various mappings.

\begin{figure}[tb]
	\centering
	\includegraphics[width=4.25in]{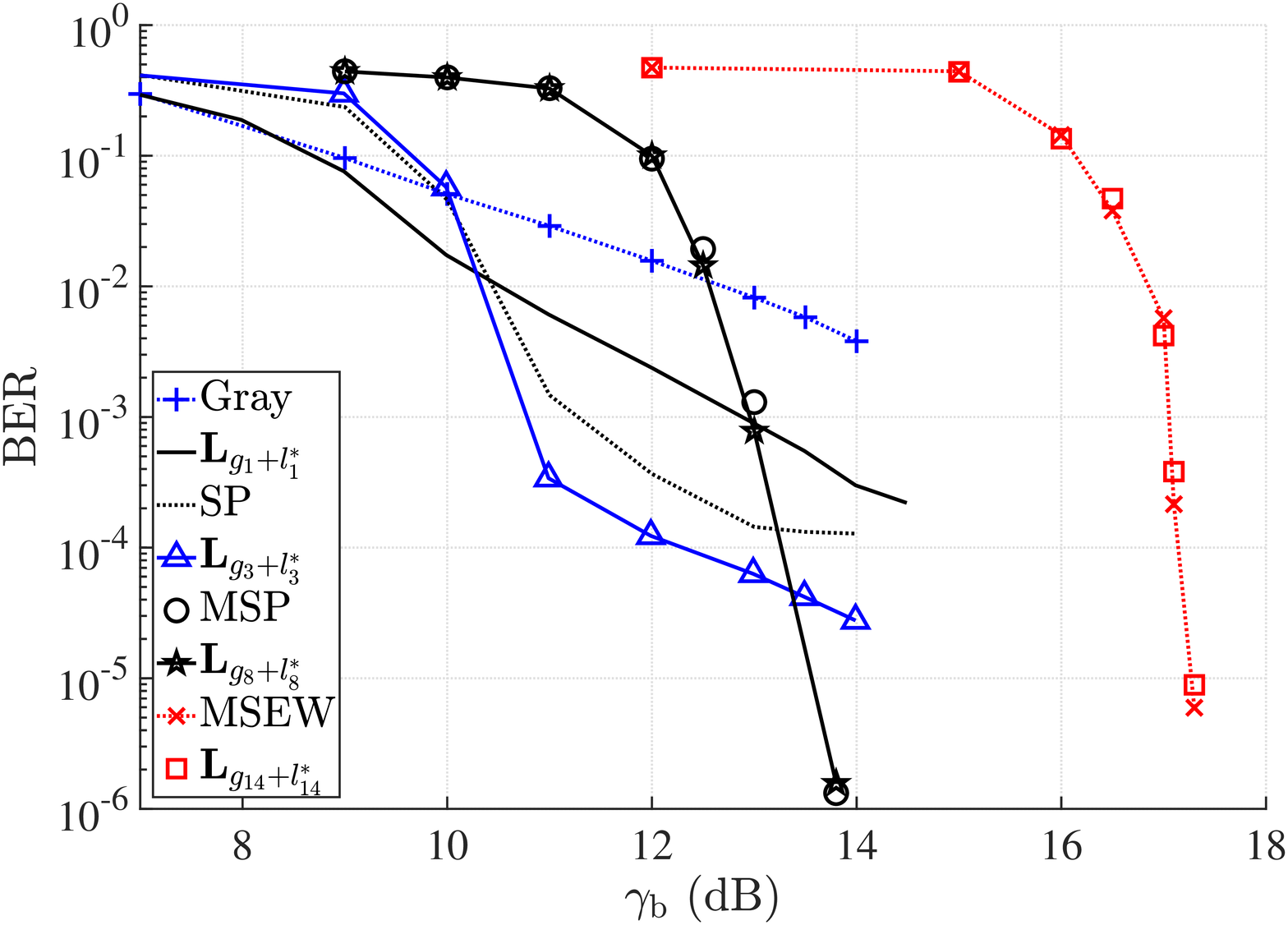}
	\caption{BER comparison of well-known mappings Gray, SP, MSP, and MSEW with relevant mappings in Table \ref{table_EX4}.}
	\label{fig_map_rev24}
\end{figure}
In general, a special mapping is not chosen by Algorithm \ref{Alg1} if there exists another mapping in the table, whose performance is not worse than the considered mapping, in both the FF and EFF modes. 
To see this, let us consider the well-known Gray, set partitioning (SP), modified set partitioning (MSP), and minimum squared Euclidean weight (MSEW) mappings, which are regarded as benchmarks in the context of traditional BICM-ID \cite{tan05twc,Tran06-Broad}. Taking a look at Table \ref{table_EX4}, one cannot find the mappings Gray and SP, whose standard decimal representations are $\mathbf{L}_{\rm Gray}=[0\ 4\ 6\ 2\ 3\ 7\ 5\ 1]$ and $\mathbf{L}_{\rm SP}=[0\ 4\ 2\ 6\ 1\ 5\ 3\ 7]$. This is due to the fact that the mappings $\mathbf{L}_{g_{1}+l_{1}^*}$ and $\mathbf{L}_{g_{3}+l_{3}^*}$, respectively, outperform the Gray and SP (so these mappings are ruled out by the algorithm). More specifically, the Gray and $\mathbf{L}_{g_{1}+l_{1}^*}$ (also the SP and $\mathbf{L}_{g_{3}+l_{3}^*}$) have the same distribution of $N_1,\cdots,N_7$ in the FF mode. Yet, the Gray mapping has $N_1=14$ in the EFF mode, which is larger than $N_1=8$ for the other mapping. Comparison of $\mathbf{L}_{g_{3}+l_{3}^*}$ and SP shows $N_1=8$ for SP and $N_1=6$ for its rival in the EFF mode. On the other hand, it is easy to confirm that mappings $\mathbf{L}_{g_{8}+l_{8}^*}$ and MSP (also $\mathbf{L}_{g_{14}+l_{14}^*}$ and MSEW) have the exact same distribution of $N_1,\cdots,N_7$ in both the FF and EFF scenarios. So, basically we expect they show the same performance at all SNRs. In other words, in terms of error rate, $\mathbf{L}_{g_{8}+l_{8}^*}$ and MSP (also $\mathbf{L}_{g_{14}+l_{14}^*}$ and MSEW) are equivalent. These facts are corroborated in Fig. \ref{fig_map_rev24}. As can be observed, the asymptotic BER of BICEM-ID when $\mathbf{L}_{g_{1}+l_{1}^*}$ is employed is significantly better than that when the Gray mapping is used (better EFF performance). Yet, both mappings have almost the same BER at low SNRs (the same FF performance). Similar arguments are valid for mappings $\mathbf{L}_{g_{1}+l_{1}^*}$ and SP. On the other hand, the BER curves, when $\mathbf{L}_{g_{8}+l_{8}^*}$ and MSP (also, when $\mathbf{L}_{g_{14}+l_{14}^*}$ and MSEW) are applied match at almost all SNRs.
	
\begin{figure}[tb]
	\centering
	\includegraphics[width=4.25in]{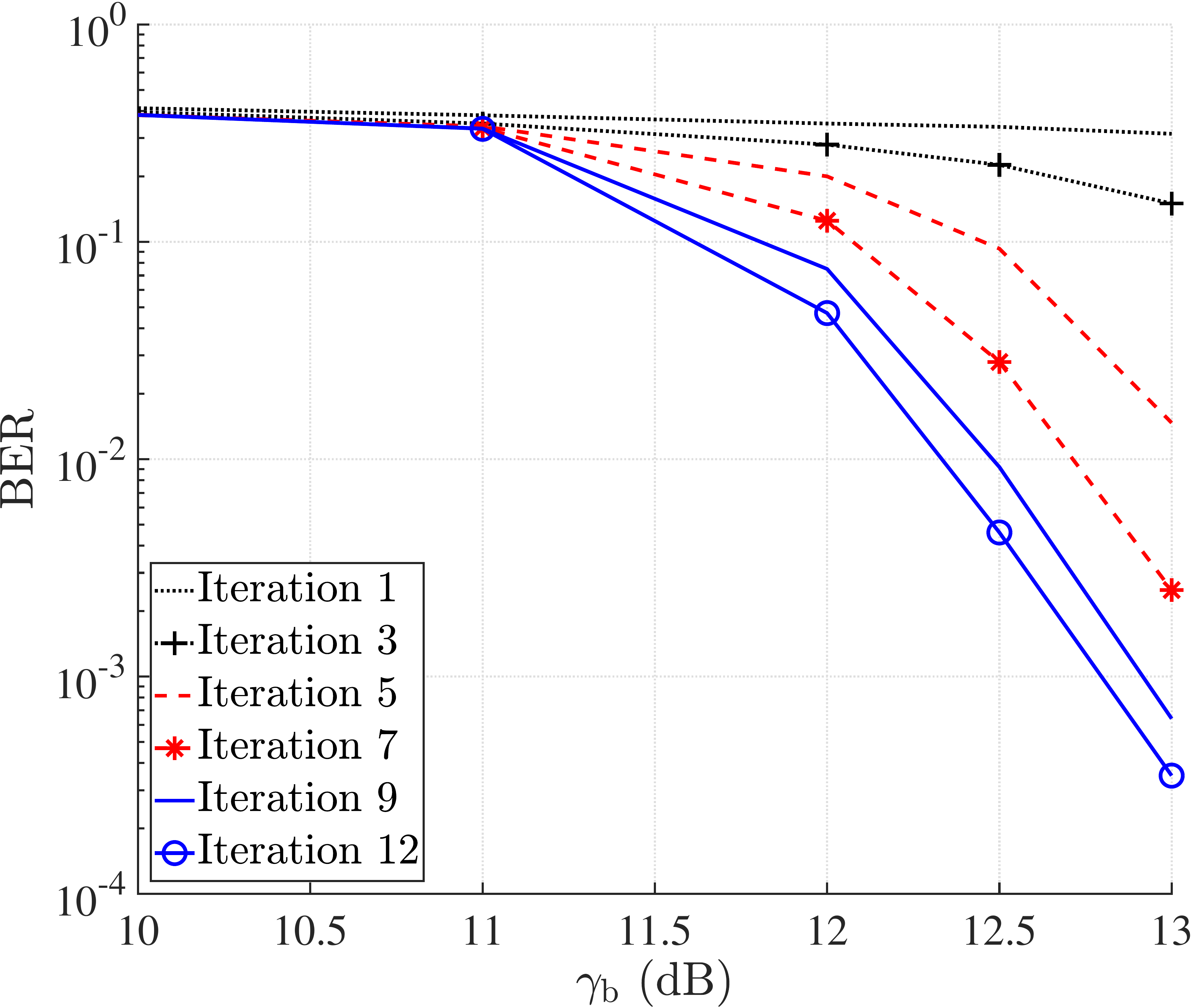}
	\caption{BER comparison of BICEM-ID for different iterations, when $\mathbf{L}_{g_8+l_8^*}$ mapping is applied.}
	\label{fig_map_comp_rew37}
\end{figure}

It is important to study the effect of iterations on the performance of BICEM-ID. Figure \ref{fig_map_comp_rew37} plots BER curves for different numbers of iterations when mapping $\mathbf{L}_{g_8+l_8^*}$ is employed. As mentioned earlier, iterations are effective (i.e., the optimal performance in the EFF mode is achievable), when the error rate in the FF mode is good enough. As such, we note that at low SNRs (up to $11.5$ dB) iterations do not help much reducing the BER. On the contrary, when SNR gets large enough ($\gamma_b$ beyond $12.5$ dB), the significant effect of iterations on reducing BER is evident.

\begin{figure}[tb]
	\centering
	\includegraphics[width=4.25in]{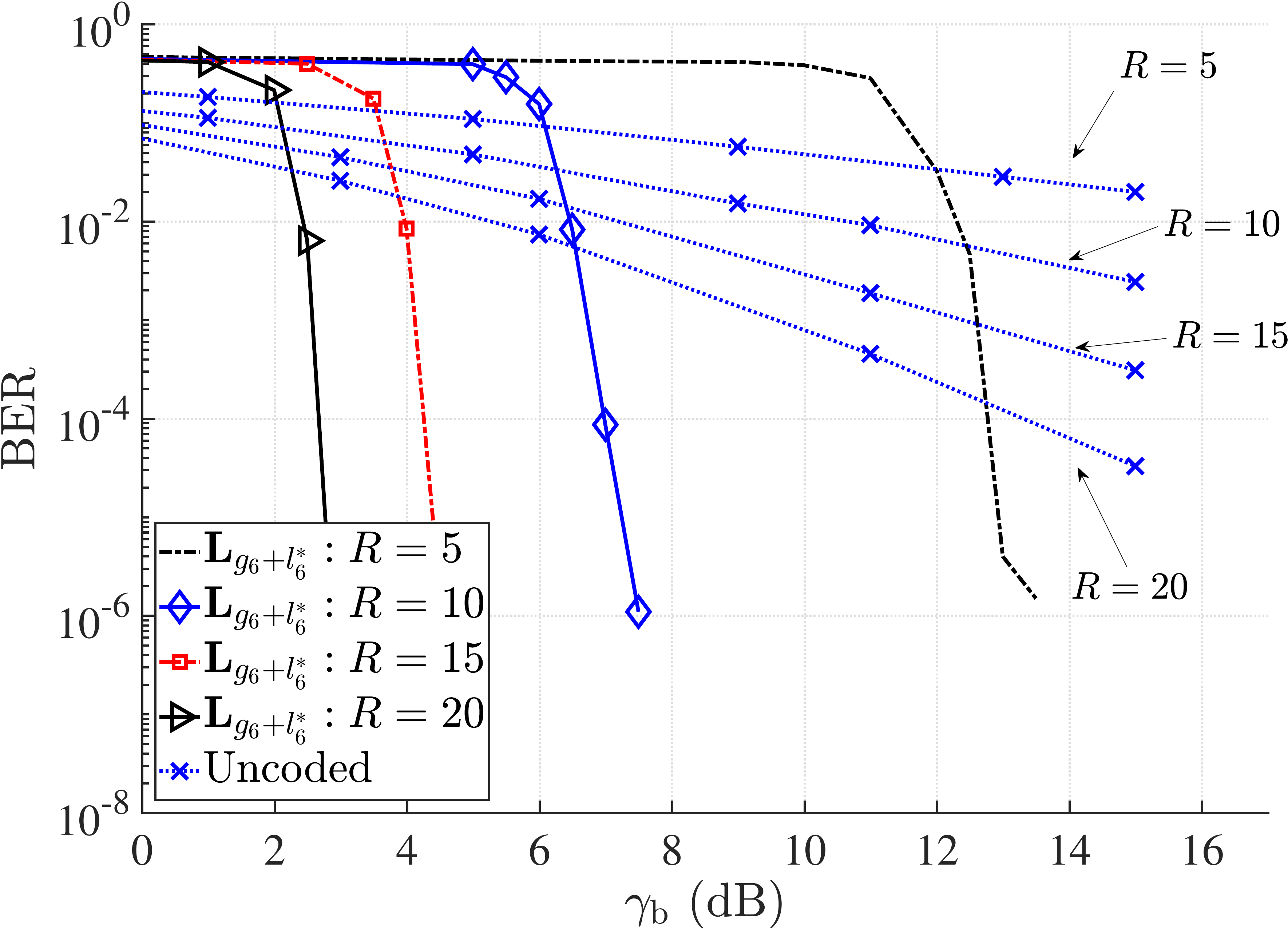}
	\caption{BERs of $8$-ary BICEM-ID with the mapping $\mathbf{L}_{g_6+l_6^*}$ in Table \ref{table_EX4} for different numbers of receive antennas.}
	\label{fig_ant_comp}
\end{figure}

It was shown in \eqref{Eq__diversity} that the system's diversity order is proportional to the number of receive antennas, $R$. Increasing the diversity order in the FF mode makes the BICEM-ID system to reach its EFF performance at lower SNRs. As a result, increasing $R$ not only increases the asymptotic slope of the BER curve, but also translates into a coding gain in the overall performance of BICEM-ID. This fact is illustrated in Fig. \ref{fig_ant_comp}, where the BER curves are plotted for the proposed $8$-ary BICEM-ID system employing the $\mathbf{L}_{g_6+l_6^*}$ mapping and when $R$ changes from $5$ to $20$. As can be seen from the figure, as $R$ increases, the performance gain of the proposed system over the uncoded one starts at lower SNRs.

\begin{figure}[tb]
	\centering
	\includegraphics[width=4.25in]{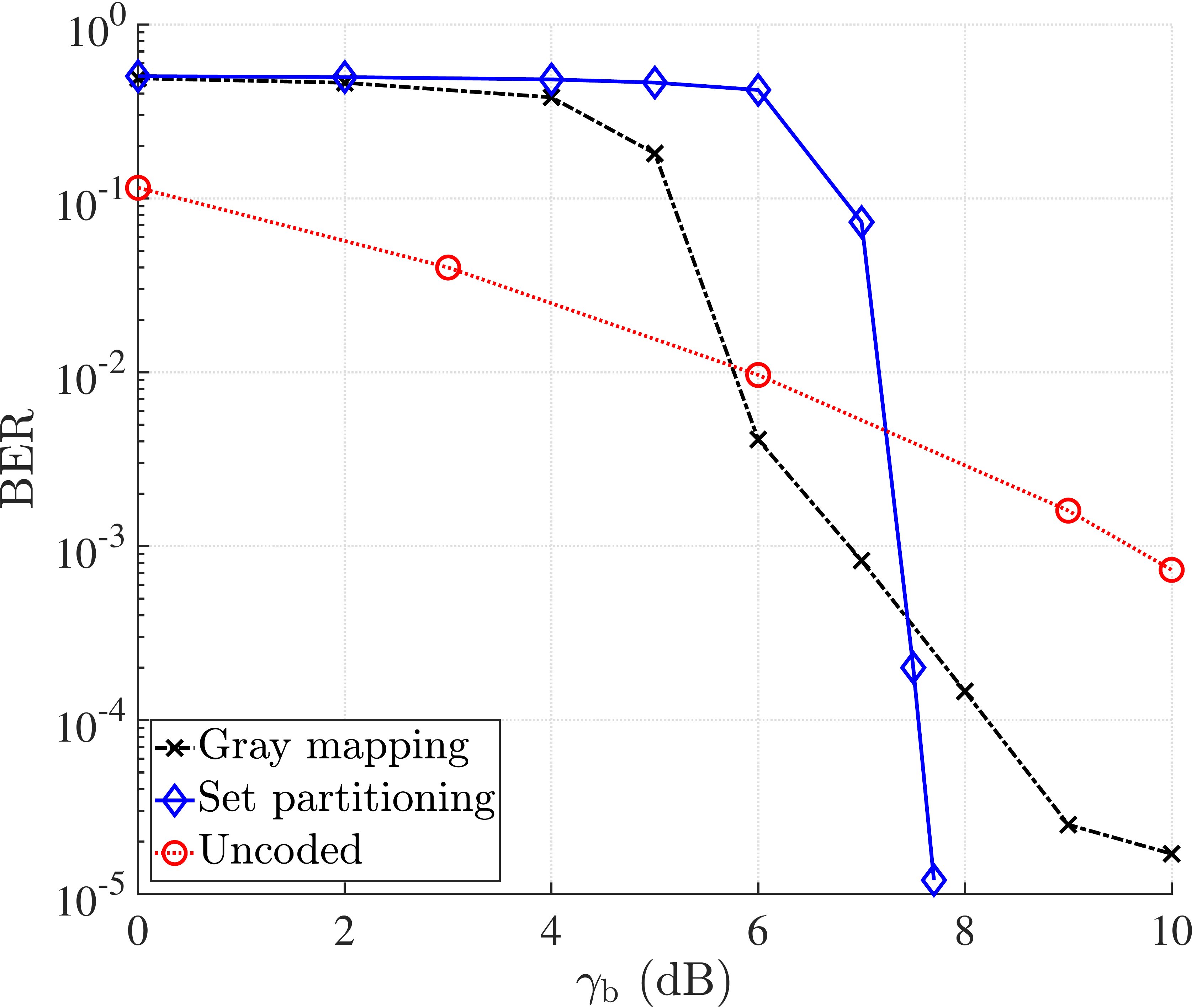}
	\caption{BERs of the $4$-ary BICEM-ID systems employing the ``natural'' and ``set-partitioning'' mappings, and uncoded OOK system (all having the same spectral efficiency of $1$ bit/channel use).}
	\label{fig_modul_comp}
\end{figure}

Finally, Fig. \ref{fig_modul_comp} compares the BERs of a $4$-ary BICEM-ID system (having spectral efficiency of $1$ bit/channel use), under two mappings, Gray and set partitioning, with mapping vectors $\textbf{L}_{\rm Gray}=[0\ 2\ 3\ 1]$ and $\textbf{L}_{\rm SP}=[0\ 2\ 1\ 3]$, respectively. These two mappings are the best ones obtained by Algorithm \ref{Alg1} for $m=2$. An implication of the BER curves in Fig. \ref{fig_modul_comp} is that the natural mapping has a better error rate in the FF mode since the curve achieves its asymptotic performance at lower SNRs. Yet, when the set partitioning mapping is applied, the asymptotic performance is better than using the natural mapping. The BERs of the coded systems are also compared and shown to be much better than the BER of the uncoded system having the same spectral efficiency.

\section{Conclusions}\label{sec-con}

A practical realization of a recently-developed non-coherent multi-level index modulation framework has been presented in this paper. The proposed system, called bit-interleaved coded EB modulation with iterative decoding (BICEM-ID), combines the technique of bit-interleaved channel coding and EB modulation to enable non-coherent detection and high spectral efficiencies. To cope with the prohibitive complexity of the maximum likelihood receiver, a suboptimal non-coherent receiver based on iterative processing has been also developed. The system's error performance has been analyzed by computing bounds on the average PEP and asymptotic BER in two scenarios: feedback-free (FF) and error-free feedback (EFF). The analysis reveals the dependence of the BER on different system parameters and leads to a design criterion of best signal mappings. Based on the obtained criterion an algorithm for finding the set of best mappings is proposed. Finally, simulation results have corroborated and highlighted the main analytical findings. 

\appendices

\section{Derivation of \eqref{Eq_PEP_Chernoffbound}}\label{appxA}
Let $X=\sum_{a=1}^{R}\sum_{v=1}^{d}\left(1-r^{q_{v}-\hat{q}_{v}}\right)| Z_{v,a} |^2$ and $C=R\ln(r)\sum_{v=1}^{d}\left(\hat{q}_{v}-q_{v}\right)$. Then according to the Chernoff bound we have
 \begin{IEEEeqnarray}{cc}
	\label{Eq_App_Chernoff}
	\Pr(\mathbf{s}\rightarrow \hat{\mathbf{s}}|\mathbf{s}) \leq \frac{M_X(t)}{\exp(tC)},
	\IEEEeqnarraynumspace
\end{IEEEeqnarray}
for every $t\geq0$, where $M_X(t)=E_X[\exp(tX)]$ is the moment generating function (MGF) of $X$, and $\Pr(\mathbf{s}\rightarrow \hat{\mathbf{s}}|\mathbf{s})$ is given by \eqref{Eq_PEP_expression2}. It is simple to see that
 \begin{IEEEeqnarray}{cc}
	\label{Eq_App_Chernoff0}
	\exp(tC)= \prod_{v=1}^d  r^{tR(\hat{q}_v-q_v)}.
	\IEEEeqnarraynumspace
\end{IEEEeqnarray}
The MGF of $X$ can be written as
\begin{IEEEeqnarray}{cc}
	\label{Eq_App_Chernoff2}
	M_{X}(t)=\prod_{v=1}^d \prod_{a=1}^{R}  M_{X_{a,v}}(t),
	\IEEEeqnarraynumspace
\end{IEEEeqnarray}
where $X_{a,v}=\left(1-r^{q_{v}-\hat{q}_{v}}\right)|Z_{v,a}|^2$, and $Z_{v,a}$ is an exponential random variable of rate $1$. It follows from the definition of the MFG that
\begin{IEEEeqnarray}{cc}
	\label{Eq_App_Chernoff3}
	M_{X_{a,v}}(t)=(1-t\lambda^{-1})^{-1}=\left(1-t(1-r^{q_{v}-\hat{q}_{v}})\right)^{-1}.
	\IEEEeqnarraynumspace
\end{IEEEeqnarray}
Substituting \eqref{Eq_App_Chernoff3} into \eqref{Eq_App_Chernoff2}, and then, substituting \eqref{Eq_App_Chernoff2} and \eqref{Eq_App_Chernoff0} into \eqref{Eq_App_Chernoff} yields the upper-bound in \eqref{Eq_PEP_Chernoffbound}.
Note that \eqref{Eq_App_Chernoff3} is well-defined only for $(1-r^{q_{v}-\hat{q}_{v}})t<1$. Since $r>1$, and $q_v-\hat{q}_v$ can take on any integer value between $-M$ and $M$, inclusive, the MFG in \eqref{Eq_App_Chernoff3} is well-defined if $-| 1-r^{M} |^{-1} \leq t\leq | 1-r^{-M} |^{-1}$. Recalling that $t\geq 0$ was the initial condition for the validity of the Chernoff bound, and taking its intersection with $-| 1-r^{M} |^{-1} \leq t\leq | 1-r^{-M} |^{-1}$, leads to the interval $0 \leq t\leq | 1-r^{-M} |^{-1}$, as given by \eqref{Eq_PEP_Chernoffbound}.

\section{Nearest Neighbors in BICEM}\label{appxB}
Consider a signal point $s^{[\ell]}\in \Psi_{b}^w$, and the set $\Psi_{\bar{b}}^w$ that includes $\frac{M+1}{2}$ signal points. We want to find a $s^{[\hat{\ell}]}\in \Psi_{\bar{b}}^w$ such that the PEP $\Pr({s^{[\ell]}}\rightarrow s^{[\hat{\ell}]}|{s^{[\ell]}})$ is maximized. Write
 \begin{IEEEeqnarray}{cc}
	\label{Eq_appA_PEP}
\!\!\! \!\! \! \!\Pr({s^{[\ell]}}\rightarrow s^{[\hat{\ell}]}|{s^{[\ell]}})&\IEEEnonumber\\
=\Pr\{\ln(f(y| &s^{[\ell]}))<\ln(f(y| s^{[\hat{\ell}]}))|\ s^{[\ell]}\},
 \IEEEeqnarraynumspace
\end{IEEEeqnarray}
where using \eqref{Eq_singleshot_levels},
 \begin{IEEEeqnarray}{cc}
	\label{Eq_appA_fy}
	\ln(f(y| &s^{[x]}))\propto -x\ln(r)-\frac{| y|^2}{N_0r^x},
	\IEEEeqnarraynumspace
\end{IEEEeqnarray}
for $x\in\{\ell,\hat{\ell}\}$. From \eqref{Eq_appA_PEP} and \eqref{Eq_appA_fy}, and using the fact that $y\sim \mathcal{C}\mathcal{N}(0,N_0r^{\ell})$, it is concluded that
 \begin{IEEEeqnarray}{cc}
	\label{Eq_appA_PEP2}
	\!\!\! \!\! \! \! \Pr({s^{[\ell]}}\rightarrow s^{[\hat{\ell}]}|{s^{[\ell]}})
	=\Pr\left\{(1-r^{\ell-\hat{\ell}})\Upsilon>(\hat{\ell}-\ell)\ln(r)\right\},
	\IEEEeqnarraynumspace
\end{IEEEeqnarray}
where $\Upsilon$ is an exponential random variable of rate $\lambda=1$. Hence,
\begin{IEEEeqnarray}{cc}
	\label{Eq_appA_PEP3}
\Pr({s^{[\ell]}}\rightarrow s^{[\hat{\ell}]}|{s^{[\ell]}})=\left\{\begin{matrix}
 	\exp\left(\varphi(\ell-\hat{\ell})\ln(r)\right)\ \ \ \ \ \ \ \ \ \hat{\ell}>\ell\\
 	1-\exp\left(\varphi(\ell-\hat{\ell})\ln(r)\right) \ \ \ \ \hat{\ell}<\ell
 \end{matrix}  \right.,
	\IEEEeqnarraynumspace
\end{IEEEeqnarray}
where $\varphi(x)=\frac{x}{1-r^x}$ is strictly increasing. Then, it is immediately concluded from \eqref{Eq_appA_PEP3} that $\Pr({s^{[\ell]}}\rightarrow s^{[\hat{\ell}]}|{s^{[\ell]}})$ is maximized, i.e., $s^{[\hat{\ell}]}$ is the ``nearest neighbor'' of ${s^{[\ell]}}$ if  $| \ell-\hat{\ell}|$ is minimized. In conclusion, the term ``nearest neighbor'' is defined based on the indices of the symbols in $\Psi$.

\section{Tightest Bound on $g_{\varrho}(d,\Psi,\xi)$}\label{appxC}

In this appendix we show that the tightest upper-bound on the average PEP $g_{\varrho}(d,\Psi,\xi)$ in \eqref{Eq_average_PEP} is obtained at $t=\frac{1}{2}$. In other words, we show that the right-hand side of \eqref{Eq_average_PEP} is minimized at $t=\frac{1}{2}$. This is equivalent to
 \begin{IEEEeqnarray}{cc}
	\label{Eq_app1_argmin}
	\argmin_{0\leq t\leq \frac{1}{1-r^{-M}}} \left\{\sum_{l=0}^{M}\sum_{w=1}^{m}\left(\frac{r^{t(l-\varrho(l,w))}}{1-t(1-r^{l-\varrho(l,w)})}\right)^R\right\}=\frac{1}{2}. \IEEEeqnarraynumspace
\end{IEEEeqnarray}
Rewrite the argument of \eqref{Eq_app1_argmin} in a more compact form as
 \begin{IEEEeqnarray}{cc}
	\label{Eq_app1_arg_compact}
e(t)=\sum_{l=0}^{M}\sum_{w=1}^{m}\vartheta^R(u_{l,w},t), \IEEEeqnarraynumspace
\end{IEEEeqnarray}
where
 \begin{IEEEeqnarray}{cc}
	\label{Eq_app1_arg_compact2}
\vartheta(u_{l,w},t)=\frac{u^t_{l,w}}{1-t(1-u_{l,w})},
\end{IEEEeqnarray}
and $u_{l,w}=r^{l-\varrho(l,w)}$. It is simple to confirm that $\vartheta(u_{l,w},t)>0$ and $\frac{\partial^2}{\partial t^2}\vartheta(u_{l,w},t)>0$, for $0\leq t\leq \frac{1}{1-r^{-M}}$. This means that $\vartheta(u_{l,w},t)$ is a strictly positive convex function. Choose the convex function $h(t)=(\max\{0,t\})^R$, and note that $\vartheta^R(u_{l,w},t)=h(\vartheta(u_{l,w},t))$. Since $\vartheta(u_{l,w},t)$ and $h(t)$ are convex functions and $h(t)$ is a non-decreasing function, $\vartheta^R(u_{l,w},t)=h(\vartheta(u_{l,w},t))$ is a convex function. Consequently, $e(t)$ in \eqref{Eq_app1_arg_compact} (which is a sum of convex functions $\vartheta^R(u_{l,w},t)$) is itself a convex function. In other words, $e(t)$ has one global minimum in the domain $0\leq t\leq \frac{1}{1-r^{-M}}$. Hence, it suffices to show that $\frac{\mathrm{d} e(t)}{\mathrm{d} t}=0$ at $t=\frac{1}{2}$. Since the (outer) sum in \eqref{Eq_app1_arg_compact} is performed on all constellation points, for any $\vartheta(u_{l,w},t)=\vartheta(r^{l-\varrho(l,w)},t)$, the term $\vartheta(u_{\varrho(l,w),w},t)=\vartheta(r^{\varrho(l,w)-l},t)=\vartheta(u_{l,w}^{-1},t)$ must also be included in the sum. In other words, \eqref{Eq_app1_arg_compact} can be rewritten as
\begin{IEEEeqnarray}{cc}
	\label{Eq_app1_arg_rewrit}
	e(t)=\sum_{l}\sum_{w=1}^{m}\vartheta^R(u_{l,w},t)+\vartheta^R(u^{-1}_{l,w},t), \IEEEeqnarraynumspace
\end{IEEEeqnarray}
where the outer sum is performed on $\frac{M+1}{2}$ points (instead of $M+1$ points). Taking the derivative of \eqref{Eq_app1_arg_rewrit} at $t=\frac{1}{2}$, we have
\begin{IEEEeqnarray}{cc}
	\label{Eq_app1_arg_deriv}
	\frac{\mathrm{d}}{\mathrm{d}t}e(t)\big|_{t=\frac{1}{2}} =R\sum_{l}\sum_{w=1}^{m}\vartheta^{R-1}\left(u_{l,w},t\right)K(u_{l,w},t)\big|_{t=\frac{1}{2}}, \IEEEeqnarraynumspace
\end{IEEEeqnarray}
where
\begin{IEEEeqnarray}{cc}
	\label{Eq_app1_k_u}
	K(u_{l,w},t)=\frac{\partial}{\partial t} \left\{\vartheta(u_{l,w},t)+\vartheta(u^{-1}_{l,w},t)\right\}.\IEEEeqnarraynumspace
\end{IEEEeqnarray}
Substituting \eqref{Eq_app1_arg_compact2} into \eqref{Eq_app1_k_u} verifies that $K(u_{l,w},t)\big|_{t=\frac{1}{2}}=0$, which makes \eqref{Eq_app1_arg_deriv} equal to zero.

\ifCLASSOPTIONcaptionsoff
\newpage
\fi


\bibliographystyle{IEEEtran}

\begin{thebibliography}{10}
	\providecommand{\url}[1]{#1}
	\csname url@samestyle\endcsname
	\providecommand{\newblock}{\relax}
	\providecommand{\bibinfo}[2]{#2}
	\providecommand{\BIBentrySTDinterwordspacing}{\spaceskip=0pt\relax}
	\providecommand{\BIBentryALTinterwordstretchfactor}{4}
	\providecommand{\BIBentryALTinterwordspacing}{\spaceskip=\fontdimen2\font plus
		\BIBentryALTinterwordstretchfactor\fontdimen3\font minus
		\fontdimen4\font\relax}
	\providecommand{\BIBforeignlanguage}[2]{{%
			\expandafter\ifx\csname l@#1\endcsname\relax
			\typeout{** WARNING: IEEEtran.bst: No hyphenation pattern has been}%
			\typeout{** loaded for the language `#1'. Using the pattern for}%
			\typeout{** the default language instead.}%
			\else
			\language=\csname l@#1\endcsname
			\fi
			#2}}
	\providecommand{\BIBdecl}{\relax}
	\BIBdecl
	
	\bibitem{Book-Proakis}
	J.~G. Proakis, \emph{Digital Communications}.\hskip 1em plus 0.5em minus
	0.4em\relax McGraw-Hill, 2001.
	
	\bibitem{Torrieri04}
	D.~Torrieri, \emph{Principles of Spread-Spectrum Communication Systems.}\hskip
	1em plus 0.5em minus 0.4em\relax New York: Springer-Verlag, 2004.
	
	\bibitem{Valenti05}
	M.~C. Valenti and S.~Cheng, ``Iterative demodulation and decoding of
	turbo-coded {M}-ary noncoherent orthogonal modulation,'' \emph{IEEE J.
		Select. Areas in Commun.}, vol.~23, no.~9, pp. 1739--1747, Sep. 2005.
	
	\bibitem{Fabregas07}
	A.~G.~I. Fabregas and A.~J. Grant, ``Capacity approaching codes for
	non-coherent orthogonal modulation,'' \emph{IEEE Trans. Wireless Commun.},
	vol.~6, no.~11, p. 4004–4013, Nov. 2007.
	
	\bibitem{Choi18IM}
	J.~Choi, ``Noncoherent {OFDM-IM} and its performance analysis,'' \emph{IEEE
		Trans. Wireless Commun.}, vol.~17, no.~1, pp. 352--360, Jan. 2018.
	
	\bibitem{Manolakos2016twc}
	A.~Manolakos, M.~Chowdhury, and A.~Goldsmith, ``Energy-based modulation for
	noncoherent massive {SIMO} systems,'' \emph{IEEE Trans. Wireless Commun.},
	vol.~15, no.~11, pp. 7831--7846, Nov. 2016.
	
	\bibitem{Chowdhury16}
	M.~Chowdhury, A.~Manolakos, and A.~Goldsmith, ``Scaling laws for noncoherent
	energy-based communications in the {SIMO} {MAC},'' \emph{IEEE Trans. Inform.
		Theory}, vol.~62, no.~4, pp. 1980--1992, Apr. 2016.
	
	\bibitem{Jing2016}
	L.~Jing, E.~D. Carvalho, P.~Popovski, and A.~O. Martinez, ``Design and
	performance analysis of noncoherent detection systems with massive receiver
	arrays,'' \emph{IEEE Trans. Signal Process.}, vol.~64, no.~19, pp.
	5000--5010, Oct. 2016.
	
	\bibitem{Cuba19}
	F.~Gómez-Cuba, M.~Chowdhury, A.~Manolakos, E.~Erkip, and A.~J. Goldsmith,
	``Capacity scaling in a non-coherent wideband massive {SIMO} block fading
	channel,'' \emph{IEEE Trans. Wireless Commun.}, vol.~18, no.~12, pp.
	5691--5704, Dec. 2019.
	
	\bibitem{Xie19}
	H.~Xie, W.~Xu, W.~Xiang, B.~Li, and R.~Wang, ``Performance of {ED}-based
	non-coherent massive {SIMO} systems in correlated {R}ayleigh fading,''
	\emph{IEEE Access}, vol.~7, pp. 14\,058--14\,069, 2019.
	
	\bibitem{Xie2020}
	H.~Xie, W.~Xu, H.~Q. Ngo, and B.~Li, ``Non-coherent massive {MIMO} systems: A
	constellation design approach,'' \emph{IEEE Trans. Wireless Commun.},
	vol.~19, no.~6, pp. 3812--3825, Jun. 2020.
	
	\bibitem{Gao2020}
	X.~Gao, J.~Zhang, H.~Chen, Z.~Dong, and B.~Vucetic, ``Energy-efficient and
	low-latency massive {SIMO} using noncoherent {ML} detection for industrial
	{I}o{T} communications,'' \emph{IEEE Internet Things J.}, vol.~6, no.~4, pp.
	6247--6261, Aug. 2019.
	
	\bibitem{Fazeli22}
	A.~Fazeli, H.~H. Nguyen, H.~D. Tuan, and H.~V. Poor, ``Non-coherent multi-level
	index modulation,'' \emph{IEEE Trans. Commun.}, vol.~70, no.~4, pp.
	2240--2255, Apr. 2022.
	
	\bibitem{Basarjul16IM}
	E.~Basar, ``Index modulation techniques for 5{G} wireless networks,''
	\emph{IEEE Commun. Mag.}, vol.~54, no.~7, pp. 168--175, Jul. 2016.
	
	\bibitem{Ishikawa18}
	N.~Ishikawa, S.~Sugiura, and L.~Hanzo, ``50 years of permutation, spatial and
	index modulation: From classic {RF} to visible light communications and data
	storage,'' \emph{IEEE Commun. Surveys and Tutorials}, vol.~20, no.~3, pp.
	1905--1938, 2018.
	
	\bibitem{Mia_Wen_2019}
	M.~Wen and et~al., ``A survey on spatial modulation in emerging wireless
	systems: Research progresses and applications,'' \emph{IEEE J. Select. Areas
		in Commun.}, vol.~37, no.~9, pp. 1949--1972, Sep. 2019.
	
	\bibitem{Mesleh2008}
	R.~Y. Mesleh, H.~Haas, S.~Sinanovic, C.~Ahn, and S.~Yun, ``Spatial
	modulation,'' \emph{IEEE Trans. Veh. Technol.}, vol.~57, no.~4, pp.
	2228--2241, Jul. 2008.
	
	\bibitem{Jeganathan2009}
	J.~Jeganathan, A.~Ghrayeb, L.~Szczecinski, and A.~Ceron, ``Space shift keying
	modulation for {MIMO} channels,'' \emph{IEEE Trans. Wireless Commun.},
	vol.~8, no.~7, pp. 3692--3703, Jul. 2009.
	
	\bibitem{Basar2013}
	E.~Basar, U.~Aygolu, E.~Panayirci, and H.~V. Poor, ``Orthogonal frequency
	division multiplexing with index modulation,'' \emph{IEEE Trans. Signal
		Process.}, vol.~61, no.~22, pp. 5536--5549, Nov. 2013.
	
	\bibitem{Chang2012}
	R.~Y. Chang, S.-J. Lin, and W.-H. Chung, ``New space shift keying modulation
	with {Hamming} code-aided constellation design,'' \emph{IEEE Commun.
		Letters}, vol.~1, no.~1, pp. 2--5, Feb. 2012.
	
	\bibitem{Basar2015}
	E.~Basar, ``Multiple-input multiple-output {OFDM} with index modulation,''
	\emph{IEEE Signal Process. Lett.}, vol.~22, no.~12, pp. 2259--2263, Dec.
	2015.
	
	\bibitem{Miao_wen2017}
	B.~Zheng, M.~Wen, E.~Basar, and F.~Chen, ``Low-complexity near-optimal detector
	for multiple-input multiple-output {OFDM} with index modulation,'' in
	\emph{Proc. IEEE Int. Conf. Commun.}, Paris, France, May, 2017, pp. 1--6.
	
	\bibitem{Sugiura2010}
	S.~Sugiura, S.~Chen, and L.~Hanzo, ``Coherent and differential spacetime shift
	keying: A dispersion matrix approach,'' \emph{IEEE Trans. Commun.}, vol.~58,
	no.~11, pp. 3219--3230, Nov. 2010.
	
	\bibitem{Xu2011siglet}
	C.~Xu, S.~Sugiura, S.~X. Ng, and L.~Hanzo, ``Reduced-complexity noncoherently
	detected differential space-time shift keying,'' \emph{IEEE Signal Process.
		Lett.}, vol.~18, no.~3, p. 153–156, Mar. 2011.
	
	\bibitem{Sugiura2011}
	S.~Sugiura, C.~Xu, S.~X. Ng, and L.~Hanzo, ``Reduced-complexity coherent versus
	non-coherent {QAM}-aided space-time shift keying,'' \emph{IEEE Trans.
		Commun.}, vol.~59, no.~11, p. 3090–3101, Nov. 2011.
	
	\bibitem{Bian13}
	Y.~Bian, M.~Wen, X.~Cheng, H.~V. Poor, and B.~Jiao, ``A differential scheme for
	spatial modulation,'' in \emph{Proc. IEEE Global Telecommun. Conf.}, Atlanta,
	GA, USA, Dec. 2013, pp. 3925--3930.
	
	\bibitem{Bian2015}
	Y.~Bian, X.~Cheng, M.~Wen, L.~Yang, H.~V. Poor, and B.~Jiao, ``Differential
	spatial modulation,'' \emph{IEEE Trans. Veh. Technol.}, vol.~64, no.~7, pp.
	3262 -- 3268, Jul. 2015.
	
	\bibitem{Ishikawa2017}
	N.~Ishikawa and S.~Sugiura, ``Rectangular differential spatial modulation for
	open-loop noncoherent massive-{MIMO} downlink,'' \emph{IEEE Trans. Wireless
		Commun.}, vol.~16, no.~3, pp. 1908 -- 1920, Mar. 2017.
	
	\bibitem{Ishikawa2018tcom}
	N.~Ishikawa, R.~Rajashekar, C.~Xu, S.~Sugiura, and L.~Hanzo, ``Differential
	space-time coding dispensing with channel estimation approaches the
	performance of its coherent counterpart in the open-loop massive {MIMO-OFDM}
	downlink,'' \emph{IEEE Trans. Commun.}, vol.~66, no.~12, p. 6190–6204, Dec.
	2018.
	
	\bibitem{Xiao2019}
	L.~Xiao, P.~Xiao, Y.~Xiao, C.~Wu, D.~Mi, and I.~A. Hemadeh, ``Rectangular
	differential {OFDM} with index modulation,'' in \emph{Proc. IEEE Veh.
		Technol. Conf.}, Kuala Lumpur, Malaysia, Apr. 2019.
	
	\bibitem{Basar17}
	E.~Basar, M.~Wen, R.~Mesleh, M.~D. Renzo, Y.~Xiao, and H.~Haas, ``Index
	modulation techniques for next-generation wireless networks,'' \emph{IEEE
		Access}, vol.~5, pp. 16\,693--16\,746, 2017.
	
	\bibitem{XU19}
	C.~Xu and et~al., ``Sixty years of coherent versus non-coherent tradeoffs and
	the road from {5G} to wireless futures,'' \emph{IEEE Access}, vol.~7, pp.
	178\,246--178\,299, 2019.
	
	\bibitem{Gopi19}
	S.~Gopi, S.~Kalyani, and L.~Hanzo, ``Coherent and non-coherent multilayer index
	modulation,'' \emph{IEEE Access}, vol.~7, pp. 79\,677--79\,693, 2019.
	
	\bibitem{Fazeli2019}
	A.~Fazeli, H.~H. Nguyen, and M.~Hanif, ``Generalized {OFDM-IM} with noncoherent
	detection,'' \emph{IEEE Trans. Wireless Commun.}, vol.~19, no.~7, pp.
	4464--4479, Jul. 2020.
	
	\bibitem{Fazeli2019Letter}
	A.~Fazeli and H.~H. Nguyen, ``Code design for non-coherent index modulation,''
	\emph{IEEE Commun. Letters}, vol.~24, no.~3, pp. 477--481, Mar. 2020.
	
	\bibitem{Nguyen2019}
	M.~Hanif and H.~H. Nguyen, ``Non-coherent index modulation in {R}ayleigh fading
	channels,'' \emph{IEEE Commun. Letters}, vol.~23, no.~7, pp. 1153 -- 1158,
	Jul. 2019.
	
	\bibitem{Caire98}
	G.~Caire, G.~Taricco, and E.~Biglieri, ``Bit-interleaved coded modulation,''
	\emph{IEEE Trans. Inform. Theory}, vol.~44, no.~3, pp. 927--946, May 1998.
	
	\bibitem{PhD_Tran08}
	N.~H. Tran, ``Exploiting diversity in wireless channels with bit-interleaved
	coded modulation and iterative decoding ({BICM-ID}),'' Ph.D. dissertation,
	University of Saskatchewan, Dec. 2007.
	
	\bibitem{Chindapol01}
	A.~Chindapol and J.~A. Ritcey, ``Design, analysis, and performance evaluation
	for {BICM-ID} with square {QAM} constellations in {R}ayleigh fading
	channels,'' \emph{IEEE J. Select. Areas in Commun.}, vol.~19, no.~5, pp.
	944--957, May 2001.
	
	\bibitem{Tran06-Broad}
	N.~H. Tran and H.~H. Nguyen, ``Signal mappings of 8-ary constellations for bit
	interleaved coded modulation with iterative decoding,'' \emph{IEEE Trans. on
		Broadcasting}, vol.~52, no.~1, pp. 92--99, Mar. 2006.
	
	\bibitem{Li02}
	X.~Li, A.~Chindapol, and J.~A. Ritcey, ``Bit-interleaved coded modulation with
	iterative decoding and 8{PSK} signaling,'' \emph{IEEE Trans. Commun.},
	vol.~50, no.~8, pp. 1250--1257, Aug. 2002.
	
	\bibitem{Wei15}
	R.~Y. Wei, W.~Y. Hsu, and J.~A. Ritcey, ``Differential encoding of 16{APSK} for
	{BICM-ID},'' in \emph{Proc. Asia-Pac. Conf. Commun. (APCC), {\rm Kyoto,
			Japan}}, Oct. 2015, pp. 637--641.
	
	\bibitem{Baeza18}
	V.~M. Baeza, A.~G. Armada, W.~Zhang, M.~El-Hajjar, and L.~Hanzo, ``A
	noncoherent multiuser large-scale {SIMO} system {Relying} on {M}-{Ary} {DPSK}
	and {BICM-ID},'' \emph{IEEE Trans. Veh. Technol.}, vol.~67, no.~2, pp.
	1809--1814, Feb. 2018.
	
	\bibitem{Benedetto97}
	S.~Benedetto, D.~Divsalar, G.~Montorsi, and F.~Pollara, ``A soft-input
	softoutput {APP} module for iterative decoding of concatenated codes,''
	\emph{IEEE Commun. Letters}, vol.~1, no.~1, pp. 22 -- 24, Jan. 1997.
	
	\bibitem{Book-Shulin}
	S.~Lin and D.~J. Costello, \emph{Error Control Coding, second edition}.\hskip
	1em plus 0.5em minus 0.4em\relax Prentice-Hall, Inc. Englewood Cliffs, New
	Jersey 07458, 2005.
	
	\bibitem{Yang21iot}
	Z.~Yang, Y.~Fang, G.~Han, and K.~M.~S. Huq, ``Spatially coupled protograph
	{LDPC}-coded hierarchical modulated {BICM-ID} systems: A promising
	transmission technique for {6G}-enabled internet of things,'' \emph{IEEE
		Internet Things J.}, vol.~8, no.~7, pp. 7907 -- 7920, Apr. 2021.
	
	\bibitem{Fang19TCOM}
	Y.~Fang, G.~Zhang, G.~Cai, F.~C.~M. Lau, P.~Chen, and G.~Han,
	``Root-protograph-based {BICM-ID}: A reliable and efficient transmission
	solution for block-fading channels,'' \emph{IEEE Trans. Commun.}, vol.~67,
	no.~9, pp. 5921--5939, Sep. 2019.
	
	\bibitem{Yang20TVT}
	Z.~Yang, Y.~Fang, G.~Zhang, F.~C.~M. Lau, S.~Mumtaz, and D.~B. da~Costa,
	``Analysis and optimization of tail-biting spatially coupled protograph
	{LDPC} codes for {BICM-ID} systems,'' \emph{IEEE Trans. Veh. Technol.},
	vol.~69, no.~1, pp. 390--404, Jan. 2020.
	
	\bibitem{tan05twc}
	J.~Tan and G.~L. Stuber, ``Analysis and design of symbol mappers for
	iteratively decoded {BICM},'' \emph{IEEE Trans. Wireless Commun.}, vol.~4,
	no.~2, pp. 662--672, Mar. 2005.
	
\end{thebibliography}

\end{document}